\documentclass[%
 reprint,superscriptaddress,
 amsmath,amssymb,prb,
]{revtex4-2}

\usepackage{graphicx}
\usepackage{dcolumn}
\usepackage{bm}
\usepackage{color}
\usepackage[version=3]{mhchem} 
\usepackage{booktabs}
\usepackage{tabularx}
\usepackage{braket} 

\newcommand{\comment}[1]{\textcolor{black}{#1}}

\begin{document}

\preprint{APS/123-QED}

\title{Magnetic order and excitations in Ce$_3$TiBi$_5$ and Ce$_3$ZrBi$_5$}\thanks{This manuscript has been authored by UT-Battelle, LLC under Contract No. DE-AC05-00OR22725 with the U.S. Department of Energy.  The United States Government retains and the publisher, by accepting the article for publication, acknowledges that the United States Government retains a non-exclusive, paid-up, irrevocable, world-wide license to publish or reproduce the published form of this manuscript, or allow others to do so, for United States Government purposes.  The Department of Energy will provide public access to these results of federally sponsored research in accordance with the DOE Public Access Plan (http://energy.gov/downloads/doe-public-access-plan).}

\author{Pyeongjae Park}
\email{park@ornl.gov}
\affiliation{Materials Science \& Technology Division, Oak Ridge National Laboratory, Oak Ridge, TN 37831, USA}

\author{Qianli Ma}
\affiliation{Neutron Scattering Division, Oak Ridge National Laboratory, Oak Ridge, Tennessee 37831, USA}

\author{Wei Tian}
\affiliation{Neutron Scattering Division, Oak Ridge National Laboratory, Oak Ridge, Tennessee 37831, USA}

\author{Stuart Calder}
\affiliation{Neutron Scattering Division, Oak Ridge National Laboratory, Oak Ridge, Tennessee 37831, USA}
\author{Matthias Frontzek}
\affiliation{Neutron Scattering Division, Oak Ridge National Laboratory, Oak Ridge, Tennessee 37831, USA}

\author{G. Sala}
\affiliation{Oak Ridge National Laboratory, Oak Ridge, TN, 37831, USA}

\author{D. Mandrus}
\affiliation{Department of Materials Science and Engineering, The University of Tennessee, Knoxville, Tennessee 37996, USA}
\affiliation{Materials Science and Technology Division, Oak Ridge National Laboratory, Oak Ridge, Tennessee 37831, USA}

\author{Shirin Mozaffari}
\affiliation{Department of Physics and Astronomy, Clemson University, Clemson, SC 29634, USA}
\affiliation{Department of Materials Science and Engineering, The University of Tennessee, Knoxville, Tennessee 37996, USA}

\author{Andrew D. Christianson}
\affiliation{Materials Science \& Technology Division, Oak Ridge National Laboratory, Oak Ridge, TN 37831, USA}

\author{Matthew B. Stone}
\email{stonemb@ornl.gov}
\affiliation{Neutron Scattering Division, Oak Ridge National Laboratory, Oak Ridge, Tennessee 37831, USA}


\begin{abstract}
\noindent The $R_{3}M\mathrm{Bi}_{5}$ rare-earth intermetallics ($R$ = rare earth, $M$ = Ti, Zr, Sc) provide a versatile platform to explore how Kondo hybridization, RKKY exchange, magnetic frustration, and broken inversion symmetry may cooperate to generate unusual magnetic behavior. We present a comprehensive neutron scattering investigation of the magnetic structure, crystal electric field (CEF), and low-energy excitations in the locally noncentrosymmetric Kondo-lattice compounds Ce$_{3}$TiBi$_{5}$ and Ce$_{3}$ZrBi$_{5}$. Powder and single-crystal neutron diffraction reveals incommensurate cycloidal antiferromagnetic order in Ce$_{3}$TiBi$_{5}$ with propagation vector $\mathbf{k} = (0,0,0.388)$ and a reduced ordered moment of $m=0.53(3)\mu_{\mathrm{B}}$. Ce$_{3}$ZrBi$_{5}$ exhibits a qualitatively similar magnetic diffraction profile, with $\mathbf{k} \simeq (0,0,0.37)$. Inelastic neutron scattering measurements resolve two clear, well-separated CEF excitations in both compounds with nearly the same profile, confirming a well-isolated Kramers doublet ground state. At low energies, a broad, quasi-elastic magnetic response is observed at $T\simeq T_{\mathrm{N}}$, whose momentum-dependence is inconsistent with that expected from conventional collective excitations of localized moments. This discrepancy, along with a Kondo temperature estimate $T_{\mathrm{K}} \sim 3-5$\,K---comparable to $T_{\mathrm{N}}$---indicates sizable Kondo hybridization, which accounts for the moment reduction and the spiral magnetic order that appears to involve the magnetic hard direction. Our results place these compounds in a regime where local inversion symmetry breaking, anisotropic CEF effects, and competing Kondo and RKKY interactions collectively give rise to unconventional magnetic order.
\end{abstract}

\maketitle

\section{Introduction}
The interplay between localized $4f$ magnetic moments and itinerant conduction electrons is central to heavy-fermion physics, where strong electronic correlations and Kondo coupling drive a variety of emergent phenomena including quantum criticality, unconventional superconductivity, and complex magnetic order~\cite{stewart1984}. Ce-based intermetallics have long served as prototypical systems to explore these themes~\cite{andres1975, steglich1979, stewart1984, petrovic2001, petrovic2001_v2}. A growing direction is the incorporation of broken inversion symmetry---either globally or locally---which permits antisymmetric spin-orbit coupling (e.g., Rashba-type terms) or odd-parity $c-f$ hybridization, thereby promoting unconventional electronic states and magnetic orders~\cite{bauer2004, fischer2011, maruyama2012, hayami2014, hayami2015, mockli2021, khim2021, nogaki2022, landaeta2022, arushi2024}.

\begin{figure}[t]
\includegraphics[width=1\columnwidth]{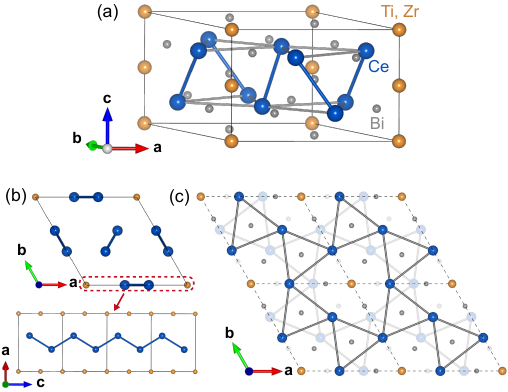} 
\caption{\label{structure} Crystal structure of Ce$_{3}$(Ti,Zr)Bi$_{5}$. (a) Crystallographic unit cell of  Ce$_{3}$(Ti,Zr)Bi$_{5}$. (b) Top and side views of the Ce zigzag chains, which form the primary motif of the magnetic sublattice. (c) In-plane connectivity between Ce sites, forming a distorted corner-shared triangular geometry. Sites below the Ce-Bi plane are shown with a lighter shading of color.} 
\end{figure}

Ce$_{3}$TiBi$_{5}$ has recently attracted attention in this context. Its hexagonal crystal structure (space group $P6_3/mcm$) hosts zigzag chains of Ce ions where inversion symmetry is locally broken at each Ce site (Fig.~\ref{structure})~\cite{motoyama2018}. This local non-centrosymmetry allows for magnetic order characterized in terms of odd-parity cluster multipoles, such as a toroidal dipole moment. As a time-reversal-odd and inversion-odd order parameter, a toroidal moment provides a natural microscopic route to magnetoelectric responses, such as current-induced magnetization~\cite{spaldin2008, hayami2014, saito2018, spaldin2019}. Indeed, early experimental studies on Ce$_{3}$TiBi$_{5}$ reported the onset of current-induced magnetization below the antiferromagnetic ordering temperature $T{_\mathrm{N}} = 5.0$\,K~\cite{motoyama2018, motoyama2020, shinozaki2020}, in qualitative agreement with theoretical proposals of a ferrotoroidal spin configuration built from two collinear antiferromagnetic sublattices along the zigzag chains, with moments pointing the chain's direction~\cite{hayami2022}. These findings identified Ce$_3$TiBi$_5$ as a rare metallic candidate potentially hosting ferro-toroidal magnetic order and associated magnetoelelectric functionality.

However, this scenario was challenged by a recent single-crystal neutron diffraction study, which reported an incommensurate single-$q$ cycloidal magnetic structure with ordering wave-vector $k$ = (0, 0, 0.386)\,(in reciprocal lattice units, r.l.u.)~\cite{gauthier2024}. In this structure---described in detail later in this paper---the Ce moments rotate within the plane of an individual zigzag chain, forming a cycloid along the crystallographic $c$-axis. Crucially, the resulting magnetic symmetry (magnetic space group P63/mcm.1'(0, 0, g)00sss) retains operations that forbid a net in-plane toroidal moment, casting doubt on interpretations of the current-induced magnetization based on a uniform in-plane toroidal order parameter.

Despite the resolution of the static magnetic structure, several aspects of Ce$_{3}$TiBi$_{5}$ merit deeper investigations, particularly from a heavy-fermion and Kondo-lattice perspective. Previous low-temperature specific heat measurements suggest a sizable Sommerfeld coefficient of $\gamma=210\,\mathrm{mJ\,K^{-2}\,mol^{-1}}$~\cite{motoyama2018}. Also, this compound exhibits unusual magnetic anisotropy behavior: while the magnetic susceptibility along the $c$-axis, which is much smaller than the $ab$-plane susceptibility, shows a conventional antiferromagnetic cusp at $T_{\mathrm{N}}$, the $ab$-plane susceptibility continues to rise as one cools below the transition~\cite{motoyama2018, motoyama2023}. Such counterintuitive behavior, in which ordering appears to involve the magnetic hard direction, could be reminiscent of hard-axis magnetism often found in several Kondo-lattice systems~\cite{krellner2008, pedder2014, hafner2019, kwasigroch2022}, where fluctuations stabilize energetically unfavorable spin configurations. These observations suggest that Kondo hybridization and crystal field effects likely play a central role in shaping the collective magnetism of Ce$_{3}$TiBi$_{5}$. Addressing these issues calls for spectroscopic probes, in particular inelastic neutron scattering (INS) to determine the crystal electric field (CEF) scheme and to search for diffuse, quasi-elastic magnetic excitations associated with possible Kondo fluctuations. Additionally, extending the material landscape to closely related compounds may provide comparative insights into how chemical substitution affects magnetic exchange, CEF splitting, and potential emergence of ferro-toroidal physics. 

In this work, we report a comprehensive neutron scattering investigation of Ce$_{3}$TiBi$_{5}$ and its isostructural analog Ce$_{3}$ZrBi$_{5}$, combining powder and single-crystal diffraction with INS. Our diffraction results confirm the previously proposed cycloidal AFM structure in Ce$_{3}$TiBi$_{5}$~\cite{gauthier2024}, which consists of two counter-rotating spirals along the $c$-axis. Neutron diffraction data on Ce$_{3}$ZrBi$_{5}$ reveal a magnetic diffraction profile similar to that of the Ti compound. INS measurements resolve two well-separated CEF excitations of the Ce$^{3+}$ $J = 5/2$ multiplet---nearly identical in both compounds---indicating a well-isolated Kramers doublet ground state manifesting pseudospin-1/2 degrees of freedom ($J_{\mathrm{eff}}=1/2$). Low-energy neutron spectroscopy at $T\simeq T_{\mathrm{N}}$ further reveals a diffuse magnetic excitation with a bandwidth of ~1\,meV that develops above $T_{\mathrm{N}}$. Comparison with spin dynamics simulations based on the established magnetic structure show qualitative disagreement, suggesting that these excitations are not conventional collective excitations of localized moments but instead of Kondo origin. Taken together, our results identify Ce$_{3}$TiBi$_{5}$ and Ce$_{3}$ZrBi$_{5}$ as Kondo-lattice systems where heavy-fermion behavior coexists with unusual magnetic order.

\section{Experimental details}
Crystals of Ce$_3$TiBi$_5$ and Ce$_3$ZrBi$_5$ were grown from a bismuth flux using an atomic ratio of Ce:Ti(Zr):Bi = 3:1:30. The Ce pieces (Ames National Laboratory), Ti (Fisher Scientific 99.95\%) or Zr slug (Fisher Scientific 99.995\%), and the Bi shot (Thermo Scientific Chemicals 99.9999\%) were loaded into an alumina crucible.  The crucible assembly was sealed within a fused silica ampule, heated to 1000 $^\circ$C over 12 h, held for 12~h, and cooled to 550 $^\circ$C at a rate of 1.2~$^\circ$C/h. At this temperature, the flux was separated from the crystals by inverting the tube and centrifuging. This process yielded needle-shaped crystals up to 1~cm in length.  These materials were found to degrade in air.  Samples were kept in inert atmospheres prior to measurements.  Samples for powder measurements were ground in a helium glove box using an agate mortar and pestle before being transferred to the respective sample container used for the measurement.

Powder neutron diffraction measurements were performed using the HB2A powder diffractometer at the High Flux Isotope Reactor (HFIR) using the Ge(113) monochromator with a calibrated wavelength of 2.408~{\AA}\cite{powderhb2a}.  The Ce$_3$TiBi$_5$ and Ce$_3$ZrBi$_5$ powders were kept in the same sample cans used for the inelastic neutron scattering measurements.  Samples were loaded into a commercial pumped liquid helium cryostat with a sample changer capable of measuring up to six samples sequentially.  Measurements were performed at $T=1.5$~K and $T=20$~K.  Data collection times were approximately 8 hours for each temperature and sample measured.  Shorter measurements were performed at additional intermediate temperatures to determine the magnetic ordering temperature.

A single crystal neutron diffraction measurement was performed using the VERITAS fixed incident energy triple axis spectrometer at the HFIR examining a crystal of Ce$_3$TiBi$_5$ mounted in the (H0L) scattering plane.  The sample was mounted to a thin aluminum plate using GE-varnish exposing the sample as little as possible to the atmosphere.  The mounted sample was placed within a pumped liquid helium cryostat.  Collimation consisted of 40'-40'-40'-80' horizontal soller collimators in place between the source and monochromator, monochromator and sample, sample and analyzer, and analyzer and detector, respectively.  A pyrolytic graphite monochromator and analyzer set for the (0 0 2) reflection (PG002) were each set to reflect 14.45 meV neutrons at the instrument.  Rocking curves at different wave-vectors were performed for temperatures between $T=1.5$ and $T=80$~K.

Single crystal neutron diffraction measurements were also performed using the WAND$^2$ (HB2C) diffractometer at the HFIR using the Ge(113) monochormator with a calibrated wavelength of 1.486~{\AA}~\cite{wandsquaredref}.  Single crystals of Ce$_3$TiBi$_5$ and Ce$_3$ZrBi$_5$  were attached to individual aluminum plates using GE-varnish.  Samples had approximate dimensions of 4~mm by 0.5~mm by 1~mm and were mounted approximately in the (HK0) scattering plane.  The long axis of the samples corresponds to the crystallographic c-axis.  Data were collected for temperatures of $T=1.5$~K and $T=10$~K with rotations of the sample about the instruments vertical axis over range of 180 degrees with 0.1 degree steps.  Data were collected for 22 seconds and 20 seconds per angular step for the Ce$_3$ZrBi$_5$ and Ce$_3$TiBi$_5$ samples respectively for each of the temperatures.

Powder inelastic neutron scattering measurements were performed using the SEQUOIA direct geometry Fermi chopper spectrometer at the Spallation Neutron Source\cite{Granroth_2010,stonespec}.  Samples were loaded into 9.5~mm diameter aluminum sample cans under an atmosphere of helium gas for thermal contact.  Sample masses were 5, 8, and 6.4 grams for the Ce$_3$TiBi$_5$, Ce$_3$ZrBi$_5$, and La$_3$TiBi$_5$ samples respectively.  Samples were co-mounted to a closed cycle refrigerator sample changer capable of holding three powder samples\cite{samplechangerpaper}.  Measurements were performed with $E_{\mathrm{i}}=4$, $30$, and $60$~meV using the high resolution Fermi chopper spinning at $120$, $240$, and $420$~Hz, respectively.  Measurements were also performed using the high flux Fermi chopper spinning at $540$~Hz set for $E_{\mathrm{i}}=540$~meV incident energy neutrons.  Measurements were performed at temperatures $T=5$, $15$, and $150$~K.  Measurements were also performed under identical conditions using an aluminum sample can with one atmosphere of helium exchange gas to be used for background subtraction purposes.

\section{Results and Discussion}

\subsection{Neutron Diffraction}

The magnetic structure investigation of Ce$_3$TiBi$_5$ and Ce$_3$ZrBi$_5$ was carried out primarily using single-crystal neutron diffraction at VERITAS and WAND$^{2}$ instruments at HFIR, ORNL. The VERITAS measurements were performed in the $(H0L)$ scattering plane, while the WAND$^2$ measurements used the $(HK0)$ geometry. These results were complemented by powder neutron diffraction measurements.

\begin{figure}
\includegraphics[scale=0.85]{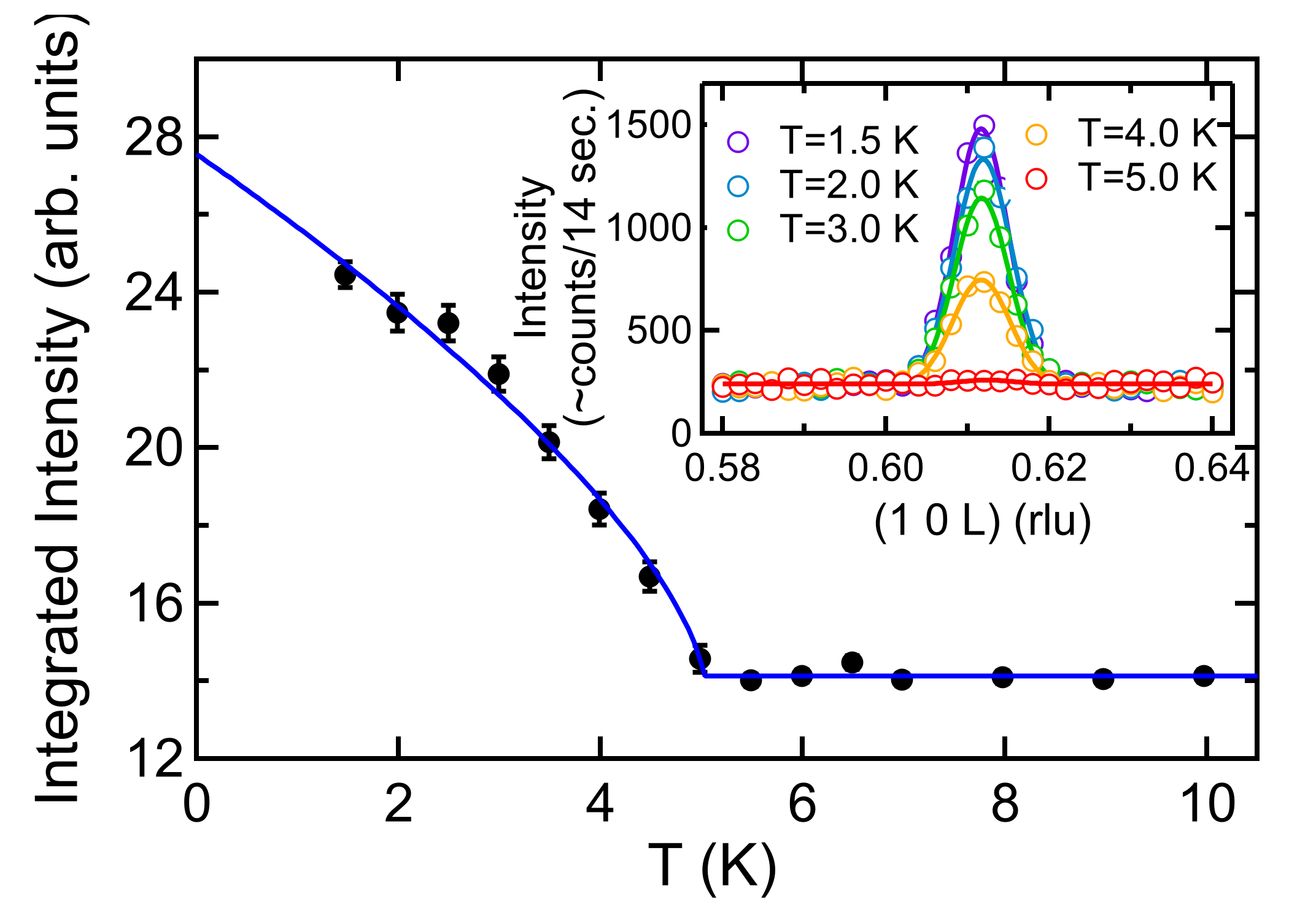} 
\caption{\label{op} Integrated scattering intensity as a function of temperature for the (1 0 0.6116) magnetic Bragg peak for Ce$_3$TiBi$_5$.  Data are based upon the integration of a fitted Gaussian peak with a constant background as shown in the inset of the figure and described in the text.  Error bars in temperature correspond to the standard deviation in the temperature for the duration of the measurement.  Solid line is a fit to a power law as described in the text. Inset: measured scattering intensity along $(1, 0, L)$ for $\hbar\omega=0$~meV from the VERITAS measurement as described in the text. Temperature listed in legend is shown to two significant digits.}
\end{figure}

Single-crystal neutron diffraction at VERITAS clearly revealed incommensurate magnetic reflections at $(H, 0, L \pm \delta)$ in Ce$_3$TiBi$_5$, with integer $H$ and $L$ and $\delta = 0.388$, at $T = 1.5$\,K. Figure~\ref{op} shows the integrated intensity of the $(1, 0, 1 - \delta)$ magnetic peak, extracted by fitting the intensity profile along the $(1, 0, L)$ direction at each temperature using a single Gaussian function plus a constant background (see inset). The temperature dependence suggesting a transition near $T_{\mathrm{N}} \sim 5$\,K confirms the magnetic origin of this reflection. Given the high data quality, the temperature dependence was further fit to a power-law expression: $I = A + B(T_{\mathrm{N}} - T)^{2\beta}$ for $T < T_{\mathrm{N}}$, and $I = A$ for $T \ge T_{\mathrm{N}}$, where $A$ is a constant background, $B$ is a scale factor, $T_{\mathrm{N}}$ is the transition temperature, and $\beta$ is the critical exponent. The fit yields $T_{\mathrm{N}} = 5.01(4)$\,K and $\beta = 0.34(3)$, in good agreement with the theoretical 3D Heisenberg critical exponent $\beta = 0.36$~\cite{chaikin1995}.

\begin{figure}
\includegraphics[scale=0.9]{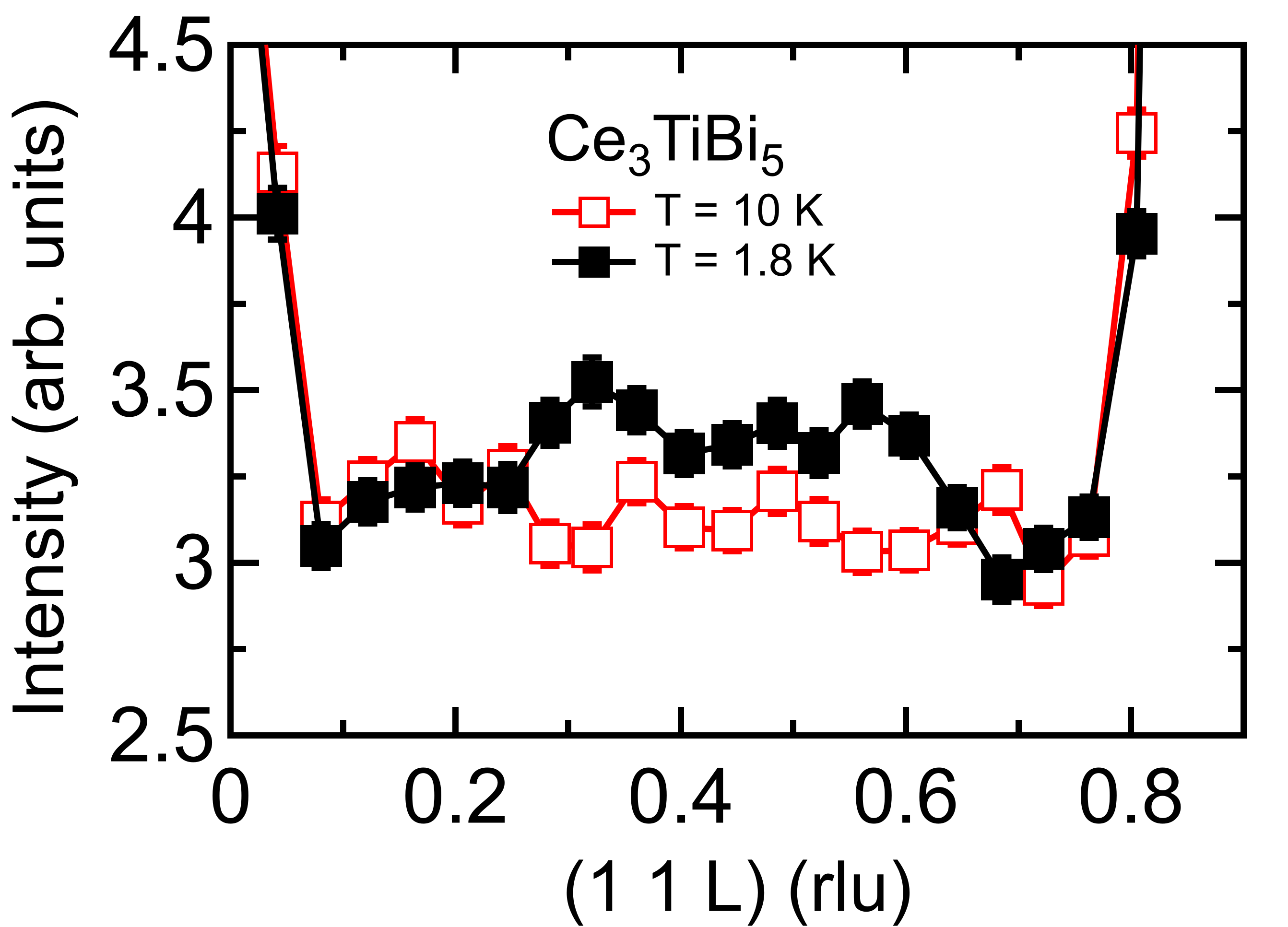} 
\caption{\label{wand2figureLdep} Scattering intensity measured along the (1, 1, $L$) direction in Ce$_3$TiBi$_5$ from WAND$^2$ at $T = 10$\,K and $T = 1.8$\,K.}
\end{figure}

\begin{figure}
\includegraphics[scale=0.90]{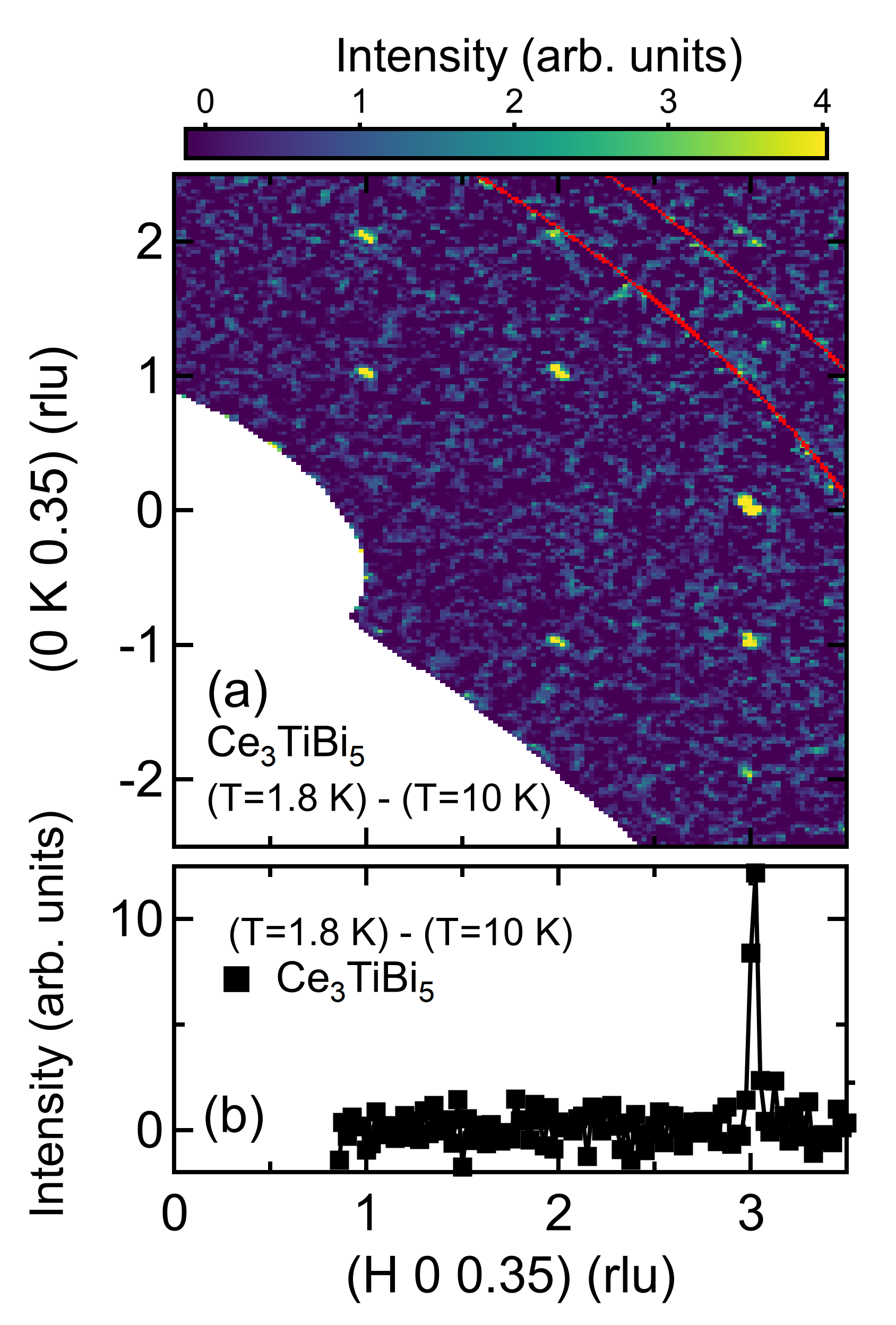} 
\caption{\label{wand2figure} In-plane magnetic reflection profile of Ce$_3$TiBi$_5$ from the WAND$^2$ measurement. (a) The difference in scattering intensity between $T = 1.8$\,K and $T = 10$\,K, integrated over $L = 0.35 \pm 0.1$,rlu. red lines mark the positions of aluminum nuclear Bragg peaks. (b) A line cut along the $(H,0,0.35)$ direction from the data in panel (a) for $K$ integrated between $\pm0.05$~rlu.}
\end{figure}

A more comprehensive diffraction survey across a wide range of momentum space was performed at WAND$^{2}$ for both Ce$_3$TiBi$_5$ and Ce$_3$ZrBi$_5$. Figure~\ref{wand2figureLdep} presents the $L$-dependent scattering intensity along the $(1, 1, L)$ direction measured above and below $T_{\mathrm{N}}$ for Ce$_3$TiBi$_5$. Magnetic reflections appear near $L = \frac{1}{3}$ and $\frac{2}{3}$, close to the incommensurate propagation vector identified in the VERITAS data (Fig.~\ref{op}). However, the intensity profiles along $L$ are notably broadened due to WAND$^2$'s vertical beam focusing, which limits the out-of-plane resolution. From these data, the incommensurate ordering vector is estimated to be $\mathbf{k} = (0, 0, \delta)$ with $\delta = 0.394$ for Ce$_3$TiBi$_5$.

\begin{figure}
\includegraphics[scale=0.7]{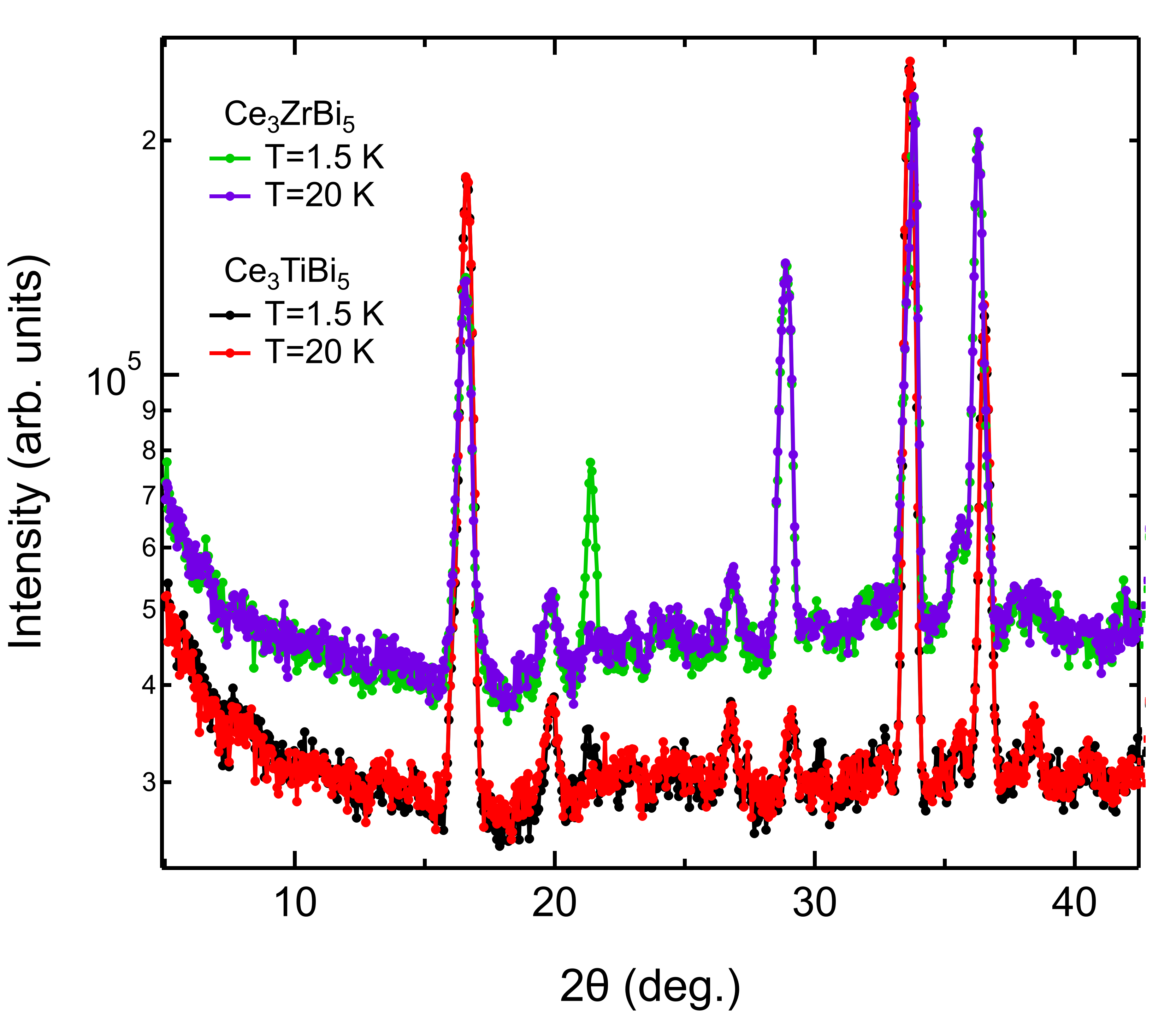} 
\caption{\label{powder_low2theta}Measured scattering intensity as a function of scattering angle, $2\theta$, for powder Ce$_3$ZrBi$_5$ and Ce$_3$TiBi$_5$ samples.  Data for the Ce$_3$ZrBi$_5$ sample are shown offset along the vertical axis for clarity and comparison.  Data are shown for $T=20$ and $T=1.5$~K measurements. The scattering angle range is chosen to illustrate the magnetic Bragg peak at $2\theta \approx 21$~degrees. Error bars in scattering intensity are smaller than symbol size.} 
\end{figure}

\begin{figure}
\includegraphics[scale=0.7]{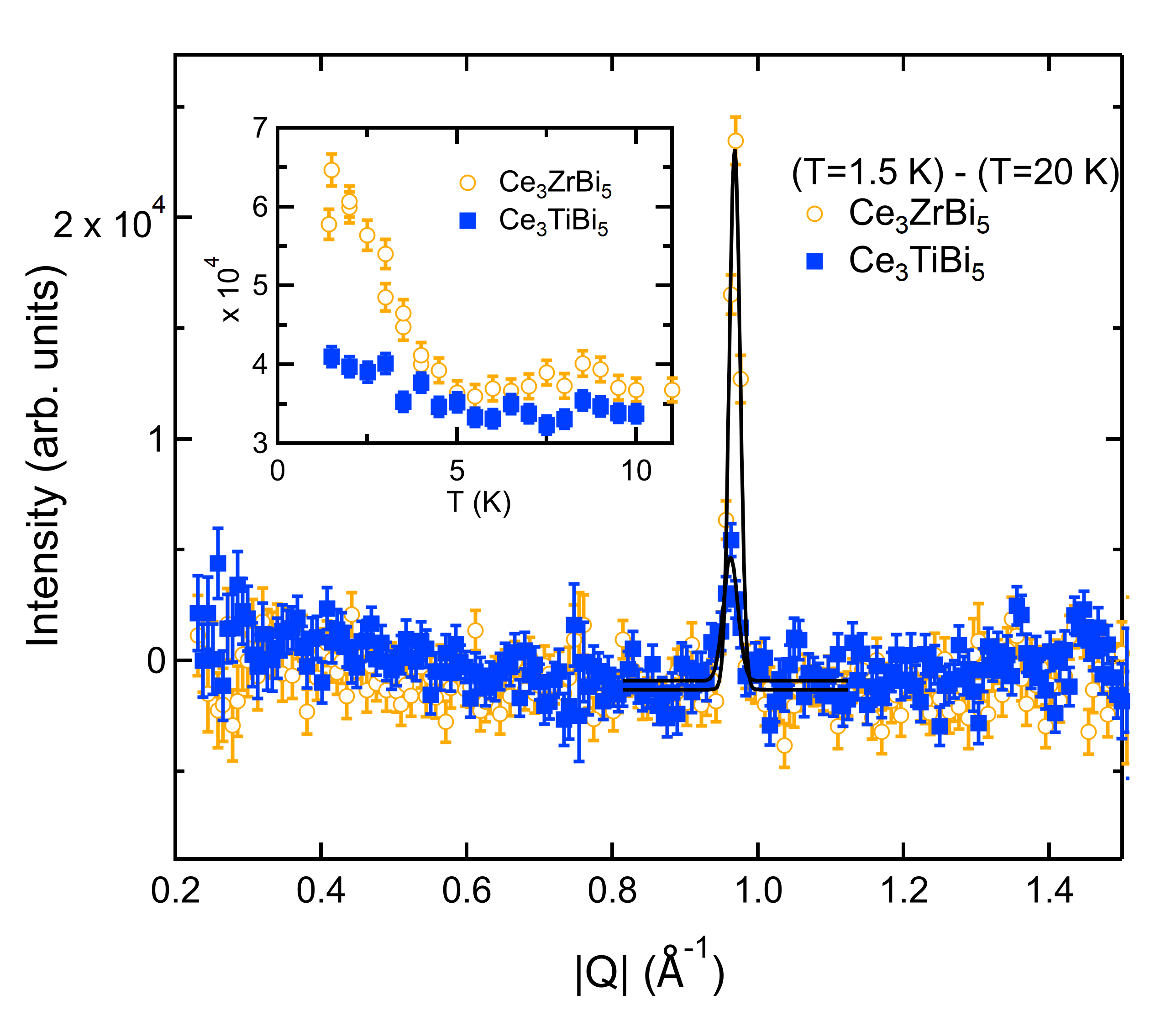} 
\caption{\label{powder_diff_vsQ}The difference in the measured scattering intensity for $T=1.5$ and $T=20$~K measurements as a function of wave-vector transfer, $|\mathbf{Q}|$, for powder Ce$_3$ZrBi$_5$ and Ce$_3$TiBi$_5$ samples.  Solid lines correspond to Gaussian fits to determine a peak position of $|\mathbf{Q}|=0.9678(3)$~\AA~for Ce$_3$ZrBi$_5$ and $|\mathbf{Q}|=0.962(1)$~\AA~for Ce$_3$TiBi$_5$.  Inset shows the temperature dependent scattering intensity (no high-temperature subtraction) for the respective samples measured with a detector of the diffractometer fixed at one scattering angle as described in the text.} 
\end{figure}

Figure~\ref{wand2figure}(a) shows an in-plane magnetic diffraction pattern of Ce$_3$TiBi$_5$, obtained by integrating the difference in scattering intensity between $T = 1.8$\,K and 10\,K over $L = 0.35 \pm 0.1$ (r.l.u.). Magnetic reflections appear at $(H, K, \delta)$ with integer $H$ and $K$, again consistent with the ordering vector $(0, 0, \delta)$. A notable feature in the diffraction pattern is the strongly suppressed intensity at the $(1, 0, \delta)$ and $(2, 0, \delta)$ positions, contrasted by the relatively strong $(3, 0, \delta)$ peak. This intensity pattern is highlighted in Fig.~\ref{wand2figure}(b), which shows a line cut along the $(H, 0, 0.35)$ direction.

Ce$_3$ZrBi$_5$ exhibits magnetic Bragg peaks at very similar positions. First, the difference in scattering intensity below and above $T_{\mathrm{N}}$, integrated over the same $L$ range of $L = 0.35 \pm 0.1$ (r.l.u.), yields a similar pattern to that of Ce$_3$TiBi$_5$ shown in Fig.~\ref{wand2figure}(a), again indicating an ordering vector $\mathbf{k} = (0, 0, \delta)$. Notably, Ce$_3$ZrBi$_5$ shares qualitatively the same profile in terms of vanishing $(1, 0, \delta)$ and $(2, 0, \delta)$ peaks. Estimation of $\delta$ based on the WAND$^{2}$ data results in $\delta = 0.363$ for Ce$_3$ZrBi$_5$, though this value may carry sizable uncertainty due to the limited vertical resolution of the instrument.

A more reliable estimate of $\delta$ for Ce$_3$ZrBi$_5$ comes from the temperature-dependent ($T=20$ and $T=1.5$~K) powder diffraction patterns shown in Figure~\ref{powder_low2theta}, focusing on the low-angle region ($2\theta<42^\circ)$. At low temperature, a clear magnetic Bragg peak appears at $2\theta \approx 21.3^\circ$ in both Ce$_3$TiBi$_5$ and Ce$_3$ZrBi$_5$. The corresponding difference spectra between the two temperatures, plotted as a function of the wave-vector $|\mathbf{Q}|$ in Fig.~\ref{powder_diff_vsQ}, further highlights this magnetic reflection at nearly identical $|\mathbf{Q}|$ values in both compounds. Gaussian fits yield peak positions of $|\mathbf{Q}| = 0.962(1)$~\AA$^{-1}$ for Ce$_3$TiBi$_5$ and $|\mathbf{Q}| = 0.9678(3)$~\AA$^{-1}$ for Ce$_3$ZrBi$_5$, which corresponds to the $(1, 0, 1-\delta)$ reflection. Estimates of $\delta$ based on these values give $\delta = 0.3967(4)$ for Ce$_3$TiBi$_5$ (close to that from Fig.~\ref{op}) and $\delta = 0.373(2)$ for Ce$_3$ZrBi$_5$.

In addition, the inset of Fig.~\ref{powder_diff_vsQ} shows measured temperature dependence of the $(1, 0, 1-\delta)$ magnetic reflection for both compounds. This temperature dependence, measured by removing the incident beam collimation and keeping the detector position fixed at $2\theta=21.2^{\circ}$ and 21.4$^{\circ}$ respectively for Ce$_3$TiBi$_5$ and Ce$_3$ZrBi$_5$, indicates an emergence of the peak in the vicinity of 5\,K, consistent with $T_{\mathrm{N}}=5\,$K for both materials~\cite{motoyama2018, motoyama2020}.

\begin{figure}
\includegraphics[width=1\columnwidth]{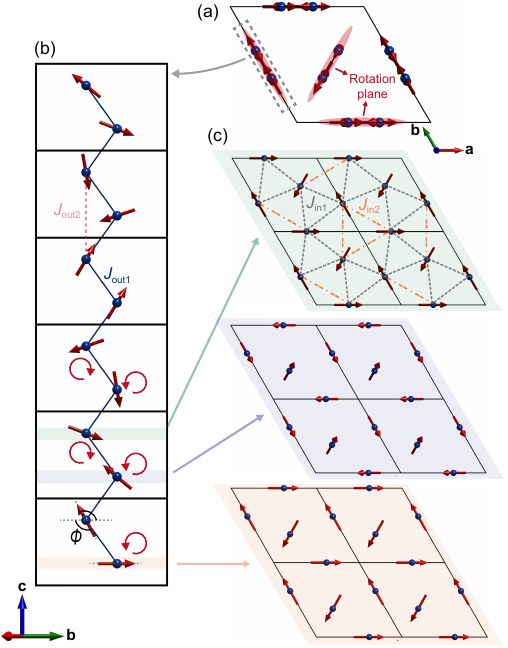} 
\caption{\label{magstr} Proposed single-$q$ spiral magnetic structure of Ce$_{3}$(Ti,Zr)Bi$_{5}$, corresponding to $\Gamma_{2}$ irreducible representation from representational analysis based on the $P6_3/mcm$ space group and the incommensurate ordering wave-vector $\mathbf{k} = (0, 0, \delta)$~\cite{gauthier2024}. (a) Schematic highlighting the spin spiral rotation plane, which is synchronized with the local zig-zag chain geometry. (b) Side view showing the two counter-rotating spiral arrangements along the $c$-axis. $\phi$ denotes the phase difference between the two counter-rotating spin spirals and is fixed at $\phi = -0.612\pi$ for the $\Gamma_2$ irreducible representation. (c) Top view of individual Ce layers, illustrating the relative phase between the three zigzag chains, which preserves six-fold rotational symmetry about the $c$-axis.}
\end{figure}

The diffraction pattern in Fig.~\ref{wand2figure}, particularly the suppression of intensity at $(1, 0, \delta)$ and $(2, 0, \delta)$ compared to the intense peak at $(3, 0, \delta)$, imposes strong constraints that enable a determination of the magnetic structure. Given that Ce$_3$TiBi$_5$ undergos a single second-order magnetic phase transition, it is reasonable to restrict the candidate magnetic structures to those arising from maximal magnetic subgroups of the parent $P6_3/mcm$ space group. As previously noted in the study of Ce$_3$TiBi$_5$~\cite{gauthier2024}, group representation analysis of the crystallographic space group with the incommensurate propagation vector $\mathbf{k} = (0, 0, \delta)$ yields six irreducible representations. Among these, only the $\Gamma_2$ and $\Gamma_5$ representations are compatible with the observed diffraction pattern in that they reproduce vanishing intensity at $(1, 0, \delta)$ and $(2, 0, \delta)$, and strong intensity at $(3, 0, \delta)$. All other representations predict significant magnetic Bragg intensities at $(1, 0, \delta)$ and $(2, 0, \delta)$---much stronger than $(3, 0, \delta)$---in contradiction with the experimental data.

From the perspective of spin configuration, both the $\Gamma_2$ and $\Gamma_5$ representations correspond to single-$q$ spiral orders where each Ce zigzag chain consists of two counter-rotating spirals. The detailed spin configuration for $\Gamma_2$ is illustrated in Fig.~\ref{magstr}. This structure retains the threefold rotational symmetry about the $c$-axis inherent in the crystallographic space group and features the following key aspects. First, the spiral rotation plane is aligned with the geometric plane of each Ce zigzag chain [Fig.~\ref{magstr}(a)]. Second, within each zigzag chain, Ce ions at $z = \frac{1}{4}c$ and $\frac{3}{4}c$ develop independent spin spirals that rotate in opposite directions [Fig.~\ref{magstr}(b)]. Third, the in-plane spin configuration across each Ce layer with corner-sharing geometry [see Fig.~\ref{structure}(c)] preserves the threefold rotational symmetry [Fig.~\ref{magstr}(c)]. Notably, this configuration is identical to the magnetic structure previously proposed for Ce$_3$TiBi$_5$ in Ref.~\cite{gauthier2024}, which has magnetic space group $P6_3/mcm.1'(0, 0, g)00sss$. Thus, as in the previous study~\cite{gauthier2024}, our diffraction results challenge the interpretation of current-induced in-plane magnetization in Ce$_3$TiBi$_5$ as arising from a ferro-toroidal spin configuration.

Magnetic structure refinement was performed via routine least-squares fitting of the observed integrated intensities using FullProf~\cite{fullprof}. This refinement was only performed for the Ti-based compound, as the measured Ce$_3$ZrBi$_5$ crystal contains at least two structural domains, complicating reliable quantitative analysis. Nonetheless, we conjecture that Ce$_3$ZrBi$_5$ adopts a similar magnetic structure, supported by the comparable incommensurate ordering vectors, similar N\`eel temperatures and magnetic susceptibility curves~\cite{motoyama2020}, and nearly identical Ce$^{3+}$ crystal field schemes, as discussed in the next section.

\begin{figure}
\includegraphics[scale=0.65]{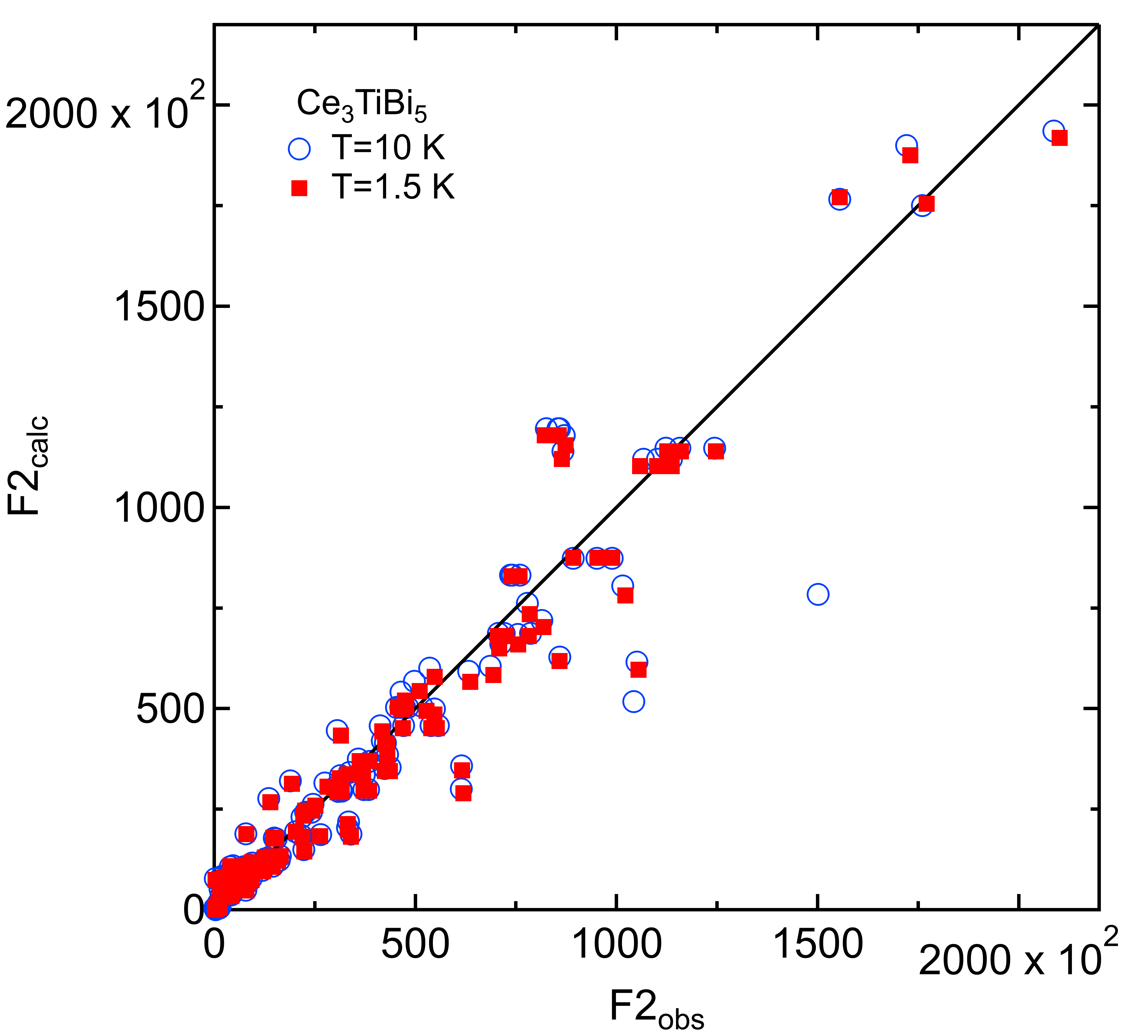} 
\caption{\label{fcalcce3tibi5nucwand2} Comparison of calculated versus measured peak intensities for the nuclear structure refinement of Ce$_3$TiBi$_5$ using WAND$^2$ data at $T=10$~K (134 peaks) and $T=1.5$~K (132 peaks). Refinement details are provided in the text. The solid black line represents the ideal case with slope 1 and zero intercept.}
\end{figure}

As an initial step, the crystal structure was refined using nuclear reflections measured at $T=10$ and 1.5\,K at WAND$^{2}$, based on the known Ce$_3$TiBi$_5$ structure~\cite{motoyama2018, gauthier2024}. The 10\,K refinement results are shown in Fig.~\ref{fcalcce3tibi5nucwand2} and Table~\ref{nuctable_CeTiBi_wand2_10K}; the 1.5\,K data yielded consistent parameters within uncertainty and are not shown separately. A similar refinement using 11 nuclear reflections from the $(H0L)$ plane at VERITAS was conducted with Lorentz factor correction, and the results are listed in Table~\ref{nuctablevertias10K}. All refinements show good agreement with the known structural model, consistent with prior reports~\cite{motoyama2018, gauthier2024}. The $c$-axis lattice constant from WAND$^2$ is not reliable due to limited vertical detector coverage but does not affect indexing or integration of magnetic reflections in the $0 < L < 1$ range. The refinement results are consistent with our complementary analysis of the crystal structure using powder neutron diffraction data ($T=20\,K$), which yielded the lattice parameters $a=9.5749(7)$ and $c=6.3954(5)$~{\AA} for Ce$_{3}$TiBi$_{5}$ and $a=9.657(3)$ and $c=6.461(2)$~{\AA} for Ce$_{3}$ZrBi$_{5}$.

\begin{table}[h]
    \centering
    \caption{Refined atomic positions for Ce$_3$TiBi$_5$ using the WAND$^{2}$ data at $T=10$~K and previously reported structure~\cite{motoyama2018, gauthier2024} having $P 63/m c m$ space group.  Values with error bars indicate refined values. The lattice constants were determined to be $a=9.589(4)$ and $c=6.80(2)$~{\AA}. The final R-factor was $R_{f}=9.83$.}
    \label{nuctable_CeTiBi_wand2_10K}
    \begin{tabular}{@{}lcccccc@{}}
        \toprule
        Atom & Element & X & Y & Z & Occ \\ \midrule
        Ce1 & Ce & 0.616(1) & 0.00000 & 0.25000 & 0.25000 \\
        Bi1 & Bi & 0.2547(7)& 0.00000 & 0.25000 & 0.25000 \\
        Bi2 & Bi & 0.33330  & 0.66670 & 0.00000 & 0.16667 \\
        Ti1 & Ti & 0.00000  & 0.00000 & 0.00000 & 0.08333 \\ \bottomrule
    \end{tabular}
\end{table}

\begin{table}[h]
    \centering
    \caption{Refined atomic positions for Ce$_3$TiBi$_5$ using the VERITAS data measured at $T=10$~K and previously reported structure~\cite{motoyama2018, gauthier2024} having $P 63/m c m$ space group.  Values with error bars indicate refined values. The lattice constants were determined to be $a=9.579(1)$ and $c=6.409(1)$~{\AA}. The final R-factor was $R_{f} =9.57$.}
    \label{nuctablevertias10K}
    \begin{tabular}{@{}lcccccc@{}}
        \toprule
        Atom & Element & X & Y & Z & Occ \\ \midrule
        Ce1 & Ce & 0.62(1) & 0.00000 & 0.25000 & 0.25000 \\
        Bi1 & Bi & 0.263(3) & 0.00000 & 0.25000 & 0.25000 \\
        Bi2 & Bi & 0.33330 & 0.66670 & 0.00000 & 0.16667 \\
        Ti1 & Ti & 0.00000 & 0.00000 & 0.00000 & 0.08333 \\ \bottomrule
    \end{tabular}
\end{table}

\begin{figure}
\includegraphics[scale=0.65]{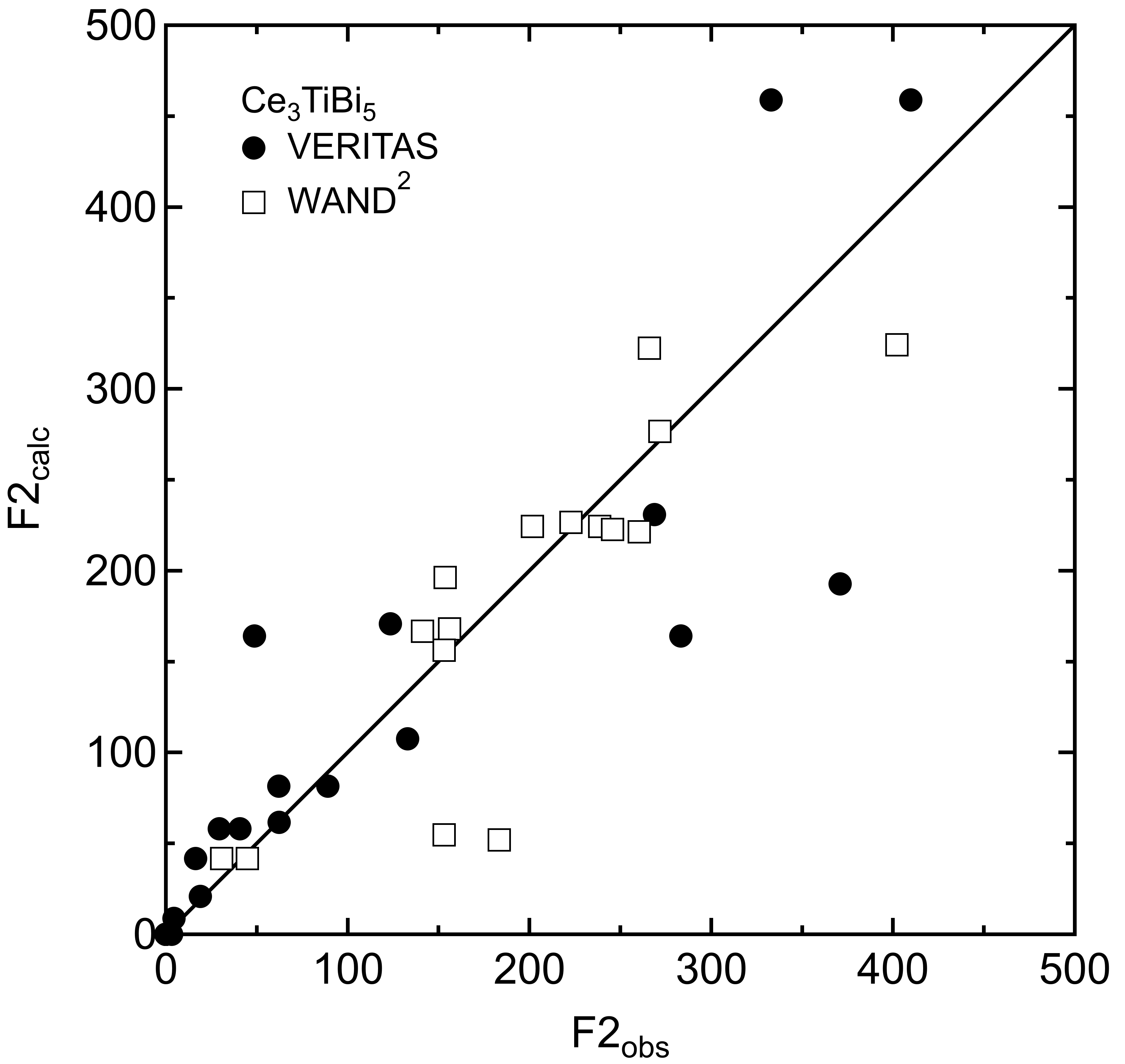} 
\caption{\label{fcalcce3tibi5magveritas} Comparison of calculated versus measured peak intensities for the magnetic structure refinement of Ce$_3$TiBi$_5$ using the VERITAS and WAND$^2$ data at $T=1.5$\,K. Refinement details and the magnetic structure model are described in the text. The solid black line represents the ideal case with slope 1 and zero intercept.}
\end{figure}

Based on the scaling factor obtained from the $T=1.5$~K nuclear refinement, we performed least-squares fits of the magnetic peak intensities. As discussed above, only the $\Gamma_{2}$ and $\Gamma_{5}$ representations qualitatively match the observed diffraction profiles. While $\Gamma_{5}$ allows slightly lower symmetry, it produces a similar spin configuration to $\Gamma_{2}$ (see Fig.~\ref{magstr}), and the subtle differences lie beyond the precision of our refinement. Thus, we adopt the $\Gamma_{2}$ model throughout, consistent with its use in~\cite{gauthier2024}. Magnetic structure refinement for Ce$_3$TiBi$_5$ was performed using Lorentz-corrected integrated intensities from 23 rocking scans collected in the $(H0L)$ scattering plane at VERITAS. Additional refinements using 16 magnetic reflections from the WAND$^2$ data were conducted for both compounds. In all cases, the $T=1.5$\,K scale factor, atomic coordinates, and anisotropic extinction parameters were fixed during the refinement.

Figure~\ref{fcalcce3tibi5magveritas} presents the refined calculated peak intensities plotted against the measured values, confirming that the $\Gamma_{2}$ magnetic structure model yields satisfactory agreement. Each refinement involved two fitting parameters: the phase difference $\phi$ between the two counter-rotating spirals [see Fig.~\ref{magstr}(b)] and the ordered magnetic moment $m$. While $\phi$ is nominally fixed at $\phi = -0.612\pi$ for the $\Gamma_{2}$ irreducible representation, it was treated as a free parameter to test robustness. From the VERITAS data on Ce$_{3}$TiBi$_{5}$, the refinement yielded $\phi = -0.46(12)\pi$ and $m = 0.56(3)\,\mu\mathrm{B}$, with an agreement factor of $R{_\mathrm{F}}=20.1$\,\%. For the WAND$^{2}$ data, we obtained $\phi = -0.64(6)\pi$ and $m = 0.50(2)\,\mu_\mathrm{B}$, with an agreement factor of $R_{\mathrm{F}} = 10.6$\,\%.

\comment{}

\subsection{Crystalline Electric Fields}
\begin{figure}
\includegraphics[scale=0.575]{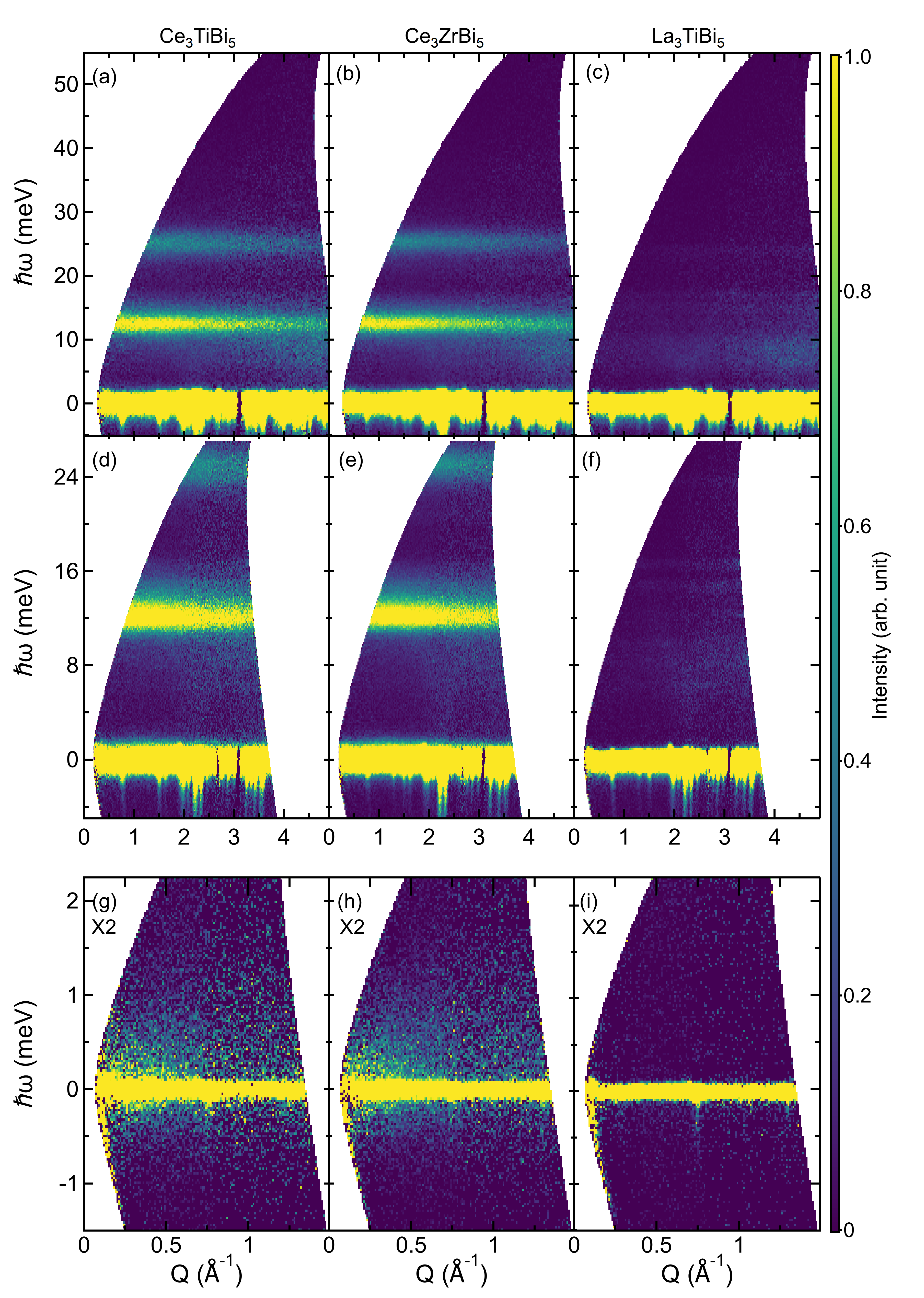} 
\caption{\label{SEQpowder} INS intensity at $T = 5$\,K as a function of energy transfer $\hbar\omega$ and magnitude of wave-vector transfer $Q$ for Ce$_3$TiBi$_{5}$ (left column), Ce$_3$ZrBi$_5$ (middle column), and La$_3$TiBi$_5$ (right column).  (a)--(c)show data collected with $E{\mathrm{i}} = 60$~meV, (d)--(f) with $E{\mathrm{i}} = 30$~meV, and (g)--(i) with $E{\mathrm{i}} = 4$~meV. The $E{\mathrm{i}} = 4$~meV data are scaled by a factor of two to be on the same intensity scale as the other measurements presented.  In all panels, the empty can background has been subtracted, and the measured intensities have been normalized by the calculated nuclear scattering cross section for each sample as described in the text.}
\label{CEF_power}
\end{figure}

Figure~\ref{SEQpowder} shows the measured neutron scattering spectra of the Ce$_3$TiBi$_5$, Ce$_3$ZrBi$_5$, and La$_3$TiBi$_5$ samples.  The empty sample can background has been subtracted from each measurement.  In addition, we divide the measured scattering intensity by the total neutron nuclear scattering cross-section of the respective sample.  Although this factor is within a value of two for the samples and their respective masses measured, this scaling allows for a better comparison of the results.  The $E_{\mathrm{i}}=60$~meV measurements, Fig.~\ref{SEQpowder}(a)-(b), indicate the presence of two non-dispersive excitation levels at energy transfers of $\hbar\omega\approx12$ and $\approx27$~meV for both samples. The higher resolution measurements for $E_{\mathrm{i}}=30$~meV shown in Fig.~\ref{SEQpowder}(d)-(e) also show the same modes in both samples.  Comparison with the La$_3$TiBi$_5$ measurement in Fig.~\ref{SEQpowder}(c) and ~\ref{SEQpowder}(f) demonstrates the absence of such modes in the non-magnetic analog, thus providing evidence that these flat modes are associated with the CEF excitations in the Ce based compounds.

\begin{figure}
\includegraphics[scale=0.7]{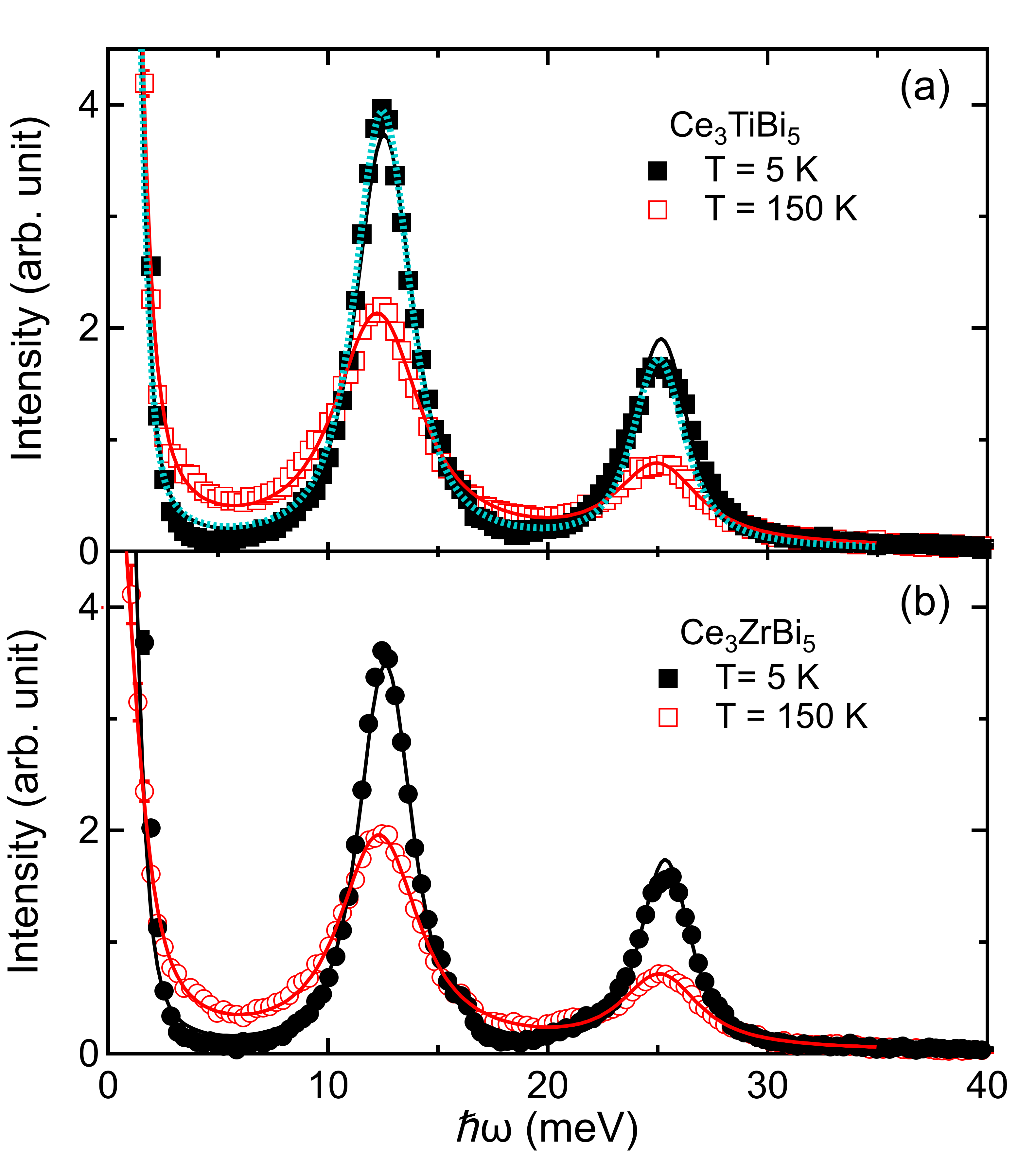} 
\caption{\label{SEQIvsE} Scattering intensity as a function of energy transfer, $\hbar\omega$, for the $E_{\mathrm{i}}=60$~meV measurements integrated over wave-vector transfers between $0 \leq Q \leq 3$ inverse Angstroms.  (a) Data for Ce$_3$TiBi$_5$ at $T=5$ and $T=150$~K. (b) Data for Ce$_3$ZrBi$_5$ at the same temperatures. All spectra have been background-subtracted and normalized as described in the text. Solid lines are fits using Lorentzian functions convolved with a fixed-width Gaussian to account for instrumental resolution at the respective energy transfer of the mode. The dashed blue curve in panel (a) represents a simulated lineshape based on the refined crystal field parameters for Ce$_3$TiBi$_5$, where the peak positions and relative integrated intensities were given by the CEF model. Note that all candidate solutions listed in Table~\ref{tab_CEFfit} yield indistinguishable lineshapes.}
\label{IvsE}
\end{figure}

Figure~\ref{SEQIvsE} presents the energy-dependent scattering intensity of the CEF excitations highlighted in Fig.~\ref{SEQpowder}(a)–(b) and (d)–(e) in more detail.  After subtracting the empty can background, the data were normalized to the calculated nuclear scattering cross section, as was done for the data in Fig.~\ref{SEQpowder}.  To remove phonon contributions, the background-subtracted and normalized La$_3$TiBi$_5$ spectrum was further subtracted from the Ce$_3$TiBi$_5$ and Ce$_3$ZrBi$_5$ data. The resulting spectra exhibit two well-defined, non-dispersive inelastic modes with similar energies and lineshapes in both Ce-based compounds.

To quantify these modes, the data in Fig.~\ref{SEQIvsE} were fit over the 0---35 meV range using a sum of three Gaussian functions (one fixed at the elastic line), each convolved with a Lorentzian to capture intrinsic broadening. The Gaussian widths were fixed to the instrumental energy resolution at each energy transfer, and a constant offset accounted for residual background. The extracted mode energies for Ce$_3$TiBi$_5$ at $T=5$\,K were $\hbar\omega_1 = 12.56(4)$\,meV and $\hbar\omega_2 = 25.16(6)$\,meV, with Lorentzian broadening of 2.4(1)\,meV. At $T=150$\,K, the modes remained nearly unchanged ($\hbar\omega_1 = 12.25(3)$\,meV, $\hbar\omega_2 = 24.98(6)$\,meV) but with a larger Lorentzian broadening of 4.19(9)\,meV. Ce$_3$ZrBi$_5$ showed comparable results: $\hbar\omega_1 = 12.62(3)$\,meV and $\hbar\omega_2 = 25.34(4)$\,meV at $T=5$\,K with a broadening 2.14(8)\,meV at $T=5$\,K, and $\hbar\omega_1 = 12.33(2)$\,meV and $\hbar\omega_2 = 25.11(4)$\,meV with a broadening 3.83(6)\,meV at $T=150$\,K. In both materials, thermal evolution of the mode positions was minimal (i.e., softening less than 0.5\,meV).

\begin{figure}
\includegraphics[width=1\columnwidth]{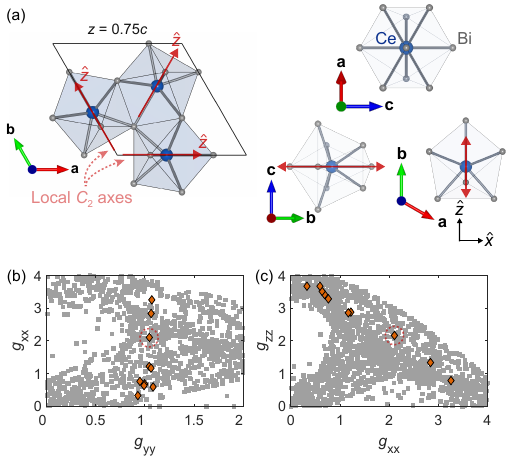} 
\caption{\label{CEF_analyis} Analysis of the single-ion CEF scheme of Ce$^{3+}$. (a) Schematic of the CeBi${8}$ polyhedron illustrating the local $C{2\upsilon}$ point symmetry and the site-dependent $C_2$ rotational axes (red arrows). The local CEF coordinate frame is defined with $\hat{z}$ along the $C_2$ axis (consistent with Eq.~\eqref{CFHam}) and $\hat{x}$ perpendicular to both $\hat{z}$ and $\hat{y} \parallel [0, 0, 1]$. (b)--(c) Distribution of $g$-tensor values calculated from $N$ candidate Stevens parameter sets $B_{n}^{m}$ sampled during the global optimization (see main text). Orange diamonds indicate solutions with minimal total loss based on both the CEF spectrum and $g$-tensor constraints (Eq.~\eqref{Eq:Loss}). The red dashed circle near the center of the figures highlights a special solution with $|g_{xx}| = |g_{zz}| \simeq 2.1$.}
\end{figure}

The extracted mode energies and intensities provide important insights into the Ce$^{3+}$ single-ion physics. The Ce$^{3+}$ ions in both Ce$_3$TiBi$_5$ and Ce$_3$ZrBi$_5$ reside at crystallographic sites with $C{_{2\upsilon}}$ point symmetry, which dictates the form of the crystal field (CEF) Hamiltonian. Yet importantly, the orientation of the local $C_2$ axis varies depending on the Ce ion’s position: it lies along the intersection of the $ab$-plane and the plane defined by the local zigzag Ce chain. As a result, the local $C_2$ axis points along one of three directions---[1, 0, 0] ($a$-axis), [0, 1, 0] ($b$-axis), or [1, 1, 0]---each case corresponding to the three distinct Ce zig-zag chains [see Fig.~\ref{CEF_analyis}(a) and ~\ref{structure}(b)].

We choose the site of Ce$^{3+}$ where the $C_2$ axis is along the crystallographic [0, 1, 0] ($b$-axis) as reference [Fig.~\ref{CEF_analyis}(a)--(b)], and the resulting CEF Hamiltonian can be described by 5 free parameters: 

\begin{eqnarray}
\mathcal{\hat H}_{\mathrm{CF}} &=& B_{2}^{0}\mathcal{\hat O}_{2}^{0} + 
B_{2}^{2}\mathcal{\hat O}_{2}^{0} + B_{4}^{0}\mathcal{\hat O}_{4}^{0} + 
B_{4}^{2}\mathcal{\hat O}_{2}^{0} + B_{4}^{4}\mathcal{\hat O}_{4}^{4},
\label{CFHam}
\end{eqnarray}
where $\mathcal{\hat O}_{n}^{m}$ represent the conventional Stevens operators\cite{stevens}, and $B_{n}^{m}$ are the Stevens parameters.

The $J = 5/2$ multiplet of Ce$^{3+}$ results in only two CEF excitation energies and one relative intensity as experimental values, providing just three constraints for five fitting parameters. However, for Ce$_{3}$TiBi$_{5}$, additional constraints can be introduced from known $g$-tensor anisotropy, extracted from field-dependent magnetization along three crystallographic directions~\cite{motoyama2023}. Specifically, the saturation behavior suggests $g{_{[100]}} \simeq 2.0$, $g_{[210]} \simeq 2.0$, and $g_{[001]} \simeq 1.0$. Including this information enables a well-constrained optimization of the Stevens parameters for Ce$_3$TiBi$_5$. Since the local CEF quantization axis $\hat{z}$ varies between different Ce sites---aligned with the direction of each zigzag chain---care must be taken when applying global $g$-tensor constraints. In the local coordinate frame, the experimentally inferred $g$-tensor translates to $g_{yy} \simeq 1.0$ and $g_{xx} + g_{zz} \simeq 4.0$.

In contrast, no such magnetization data are available for Ce$_3$ZrBi$_5$, making the fit under-constrained and leading to a broad family of compatible solutions. As a result, a quantitative CEF analysis was carried out only for Ce$_3$TiBi$_5$. Nevertheless, it is noteworthy that the observed CEF excitation energies and relative intensities are nearly identical between the two compounds [Figs.~\ref{CEF_power}(a)–(b) and (d)–(e)], suggesting that the single-ion CEF scheme is likely very similar. This is reasonable given that the local CEF environment is dominated by Bi atoms, whose atomic positions are very similar in both materials.

The fitting was performed by globally optimizing a custom-defined loss function using CrysFieldExplorer \cite{MaCEF}:

\begin{eqnarray}
L_{\mathrm{tot}} &=& L_{\mathrm{Spectrum}} + L_{\mathrm{g}},
\label{Eq:Loss}
\end{eqnarray}

\noindent where $L_{\mathrm{Spectrum}}$ is a characteristic-spectrum loss function defined as $L_{Spectrum}=L_E+L_{Intensity}$ \cite{MaCEF}. $L_{\mathrm{g}}$ is the loss function of the diagonal components of the $g$-tensor matrix. Their explicit functional expressions are:

\begin{eqnarray}
&L_E = \log_{10}\left(\sum_i^n det\{(E_{\mathrm{exp}}[i]+E_{\mathrm{cal}}[0])\mathrm{I} -H_{\mathrm{CF}} \}^2\right), \\
&L_{\mathrm{Intensity}} = \dfrac{\sqrt{\sum_i^n ({I_{\mathrm{exp}}[i]-I_{\mathrm{calc}}}[i])^2}}{\sqrt{(\sum{I_{\mathrm{exp}}}[i])^2}}, \\
&L_{\mathrm{g}} = \dfrac{\sqrt{({\mathrm{g}_{\mathrm{exp}}-\mathrm{g}_{\mathrm{calc}}})^2}}{\sqrt{({\mathrm{g}_{\mathrm{exp}}})^2}}.
\label{Eq:Loss2}
\end{eqnarray}
where the summation of $n$ iterates from $1$ to the $n$-th crystal field levels of the Ce$^{3+}$. $E_\mathrm{exp}[i]$ and $I_\mathrm{exp}[i]$ represent the experimentally observed crystal field excitation energy levels and their relative intensities observed in Fig.~\ref{CEF_power}. Their exact values are obtained by fitting Lorentzian functions convolved with a Gaussian of fixed width shown as solid black lines in Fig.~\ref{IvsE}. To perform a general sampling of the 5-dimensional CEF parameter phase space, 1920 fitting processes were performed, each time with a set of random starting parameters. The optimization routine in CrysFieldExplorer initially focuses on minimizing the $L_{Spectrum}$ to best match with the neutron spectra data, then it introduces the constraints on the $g$-tensor as described above.

Figure~\ref{CEF_analyis}(b) and (c) show the resulting distribution of solutions by visualizing the correlations between $g_{xx}$ vs $g_{yy}$ and $g_{zz}$ vs $g_{xx}$. The solid gray data points represent the general results that match neutron spectra very well. The highlighted solid orange diamond data points represent the down-selected 10 candidate solutions after introducing the constraints on $g$-tensor components. The 10 candidate solutions broadly suggest three possible scenarios of single-ion anisotropy: i) $g_{xx} \simeq$ 3.05(20), $g_{yy} \simeq$ 1.07(1), $g_{zz} \simeq$ 1.1(3) (2 solutions), ii) $g_{xx} \simeq$ 2.1, $g_{yy} \simeq$ 1.0, $g_{zz} \simeq$ 2.1 (1 solution), and iii) $g_{xx} \simeq$ 0.8(4), $g_{yy} \simeq$ 1.0(1), $g_{zz} \simeq$ 3.5(4) (7 solutions). Table~\ref{tab_CEFfit} summarizes the CEF parameters (in meV) and the ground-state doublet wavefunctions for the representative solution of each scenario. The second solution---highlighted in panels (b) and (c)---corresponds to a special case where $g_{xx} \simeq g_{zz}$. The dashed blue curve in Fig.~\ref{SEQIvsE}(a) is a line-shape based upon this highlighted solution, indicating good agreement with the measurement. 

{\renewcommand{\arraystretch}{1.1}
\begin{table}[!t]

\caption{Optimized Stevens coefficients and ground-state doublet wavefunctions for Ce$_{3}$TiBi$_{5}$. Table shows the fitted Stevens operator coefficients ($B_{n}^{m}$, in meV) and the corresponding ground-state doublet wavefunctions for the three representative solutions discussed in the main text. The local quantization axis $\hat{z}$ used in the analysis is defined in Fig.~\ref{CEF_analyis}(a). Although additional basis states such as $\ket{\pm 3/2}$ and $\ket{\mp 1/2}$ are symmetry-allowed and appear with small finite amplitudes in some solutions, their coefficients are negligible (on the order of 0.01) and are therefore omitted for clarity.}

\begin{ruledtabular}
\begin{tabular}{cccc}
 & Solution (i) & Solution (ii) & Solution (iii) \\ 
 \midrule
$B_{2}^{0}$\,(meV) & -0.6566 & -0.982 & -0.5703  \\ 
$B_{2}^{2}$\,(meV) & -1.306 & -0.0475 & 1.516 \\ 
$B_{4}^{0}$\,(meV) & 0.0086 & 0.030 & 0.0052  \\ 
$B_{4}^{2}$\,(meV) & 0.121 & -0.141 & 0.225  \\ 
$B_{4}^{4}$\,(meV) & -0.2685 & 0.1954 & 0.0193 \\
\midrule \midrule
\multicolumn{4}{c}{Ground state doublet} \\
\midrule
\multicolumn{4}{c}{(i) $0.726\ket{\pm5/2}-0.622\ket{\mp3/2}+0.295\ket{\pm1/2}$} \\
\multicolumn{4}{c}{(ii) $0.819\ket{\pm5/2}-0.537\ket{\mp3/2}+0.204\ket{\pm1/2}$} \\
\multicolumn{4}{c}{(iii) $0.843\ket{\pm5/2}+0.036\ket{\mp3/2}-0.537\ket{\pm1/2}$} \\
\midrule \midrule
\multicolumn{4}{c}{$g$-tensor} \\
\midrule
\multicolumn{4}{c}{(i) $g_{xx}=2.84$ $g_{yy}=1.06$ $g_{zz}=1.34$} \\
\multicolumn{4}{c}{(ii) $g_{xx}=2.11$ $g_{yy}=1.05$ $g_{zz}=2.17$} \\
\multicolumn{4}{c}{(iii) $g_{xx}=0.76$ $g_{yy}=0.95$ $g_{zz}=3.29$} \\
\end{tabular}
\end{ruledtabular}
\label{tab_CEFfit}
\end{table}


Finally, the INS spectra using epithermal neutrons reveal high-energy excitations likely associated with inter-multiplet transitions of Ce$^{3+}$. Figure~\ref{SEQpowderhighE} presents the energy-dependent scattering intensity from the $E_{\mathrm{i}}=540$\,meV measurements.  No background subtraction or cross-section dependent normalization has been applied to these data. Due to the reduced neutron flux at such high incident energies, we also display the empty can scattering for reference alongside the sample data. At low momentum transfers [Fig.~\ref{SEQpowderhighE}(a)], a broad mode near 300\,meV is observed in both Ce$_3$TiBi$_5$ and Ce$_3$ZrBi$_5$. At higher $|Q|$ [Fig.~\ref{SEQpowderhighE}(b)], the background increases likely due to multiphonon or multiple scattering processes, yet  the 300\,meV feature in Ce$_3$ZrBi$_5$ remains well discernible.

The appearance of this mode in both compounds, along with its $Q$-dependence, supports its identification as a single-ion excitation of Ce$^{3+}$. It likely corresponds to the spin-orbit-induced transition from the $J = 5/2$ ground-state multiplet to the $J = 7/2$ excited-state manifold, typically found near $\sim$280\,meV~\cite{jensen1991}, with the energy splitting predominantly determined by the spin-orbit coupling strength of Ce$^{3+}$. Solid lines in Fig.~\ref{SEQpowderhighE} represent independent Gaussian fits with fixed widths based on the calculated energy resolution for the nominal range of energy transfers of the apparent excitation. Enumerating this higher energy peak to be $\hbar\omega_3$, the fitted peak centers are $\hbar\omega_3=299(2)$\,meV for Ce$_3$TiBi$_5$ and 297(1)\,meV for Ce$_3$ZrBi$_5$.

\begin{figure}
\includegraphics[scale=0.75]{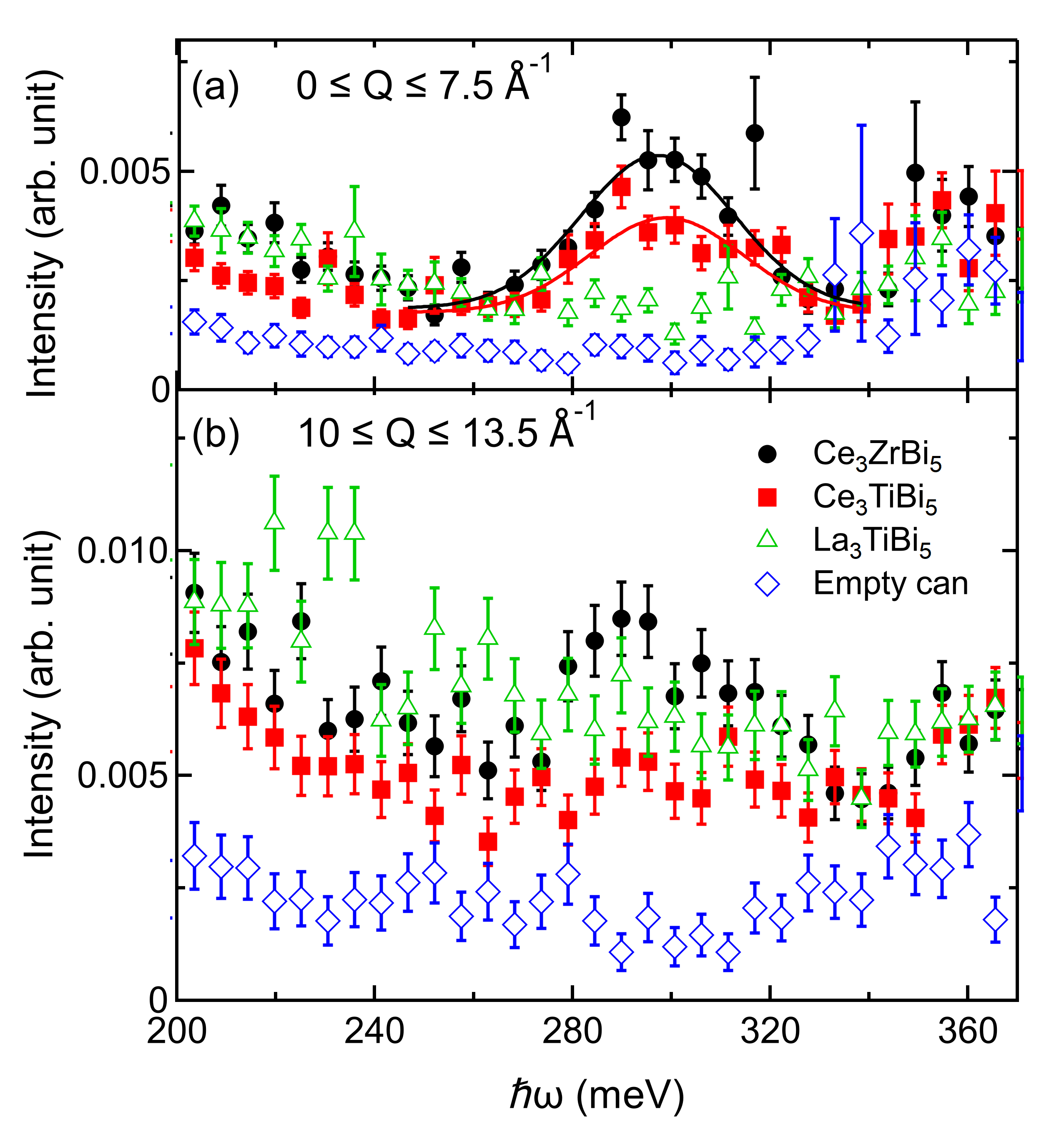} 
\caption{\label{SEQpowderhighE} $T=5$~K scattering intensity as a function of energy transfer, $\hbar\omega$, from $E_{\mathrm{i}} = 540$\,meV measurements, shown for two momentum transfer ranges: (a) $0 \leq Q \leq 7.5$\,\AA$^{-1}$ and (b) $10 \leq Q \leq 13.5$\,\AA$^{-1}$. The plotted energy window spans $200 \leq \hbar\omega \leq 360$\,meV. No background subtraction or sample-dependent scaling has been applied. Solid lines represent Gaussian fits with fixed width over the specified energy range, as described in the text.}
\end{figure}

\subsection{Low-energy magnetic excitations}
In addition to thermal neutron spectroscopy, which revealed crystal field excitations, low-$E_{i}$ measurements revealed an additional magnetic excitation at sub-meV energy scales. Figures.~\ref{SEQpowder}(g)--(i) show the $E_{\mathrm{i}}=4$\,meV results at $T=5$\,K, near the $T_{\mathrm{N}}$ of both compounds. Broad magnetic excitations are clearly visible in the Ce-based samples, but are absent in the non-magnetic La analog, suggesting a magnetic origin. These excitations may arise from either Kondo-related spin fluctuations or collective excitations of localized Ce$^{3+}$ moments in the paramagnetic phase.

To evaluate the latter scenario, we performed simulations of the paramagnetic excitation spectrum based on stochastic Landau–Lifshitz dynamics. This semiclassical approach augmented by quantum corrections has recently been shown to successfully describe energy- and momentum-resolved spectra in the paramagnetic regime, even for systems with quantum spins (which is the case of Ce$^{3+}$ with $J_{\mathrm{eff}} = 1/2$)~\cite{dahlbom2024, park2024_1, park2024_2, dahlbom2024_2, kim2025}. We used the Su(n)ny software package for the simulations~\cite{dahlbom2025sunny}, and the details of the simulation protocol can be found in Refs.~\cite{dahlbom2024, park2024_1}.

For these simulations, we constructed a minimal phenomenological spin Hamiltonian that captures the essential features of the observed magnetic structure. The model includes four exchange interactions, labeled in Fig.~\ref{magstr} as $J_{\mathrm{out1}}$, $J_{\mathrm{out2}}$, $J_{\mathrm{in1}}$, and $J_{\mathrm{in2}}$, with "out" and "in" denoting out-of-plane and in-plane interactions, respectively. Primarily, the out-of-plane interactions $J_{\mathrm{out1}}$ and $J_{\mathrm{out2}}$ drive the observed incommensurate spin modulation along the $c$-axis. This modulation can be naturally realized through frustration between $J_{\mathrm{out1}}$ and the next-nearest-neighbor coupling $J_{\mathrm{out2}}$, a scenario commonly encountered in zigzag chain systems [see Fig.~\ref{magstr}(b)]. The pitch of the resulting incommensurate modulation is directly determined by the ratio $J_{\mathrm{out2}} / J_{\mathrm{out1}}$~\cite{blundell}:

\begin{equation}
    \cos{\theta} = -\frac{J_{\mathrm{out1}}}{4J_{\mathrm{out2}}},
\end{equation}
where $\theta$ denotes the pitch angle of the spin spiral—that is, the rotation angle between nearest-neighbor (NN) spins. The experimentally determined magnetic propagation vector $k = (0, 0, 0.388)$ (r.l.u.) for Ce$_{3}$TiBi$_{5}$ implies $J_{\mathrm{out1}}<0$ (ferromagnetic) and $J_{\mathrm{out2}}=-0.725J_{\mathrm{out1}}$ (i.e., $J_{\mathrm{out2}}$ is antiferromagnetic). In addition, another key structural feature of the magnetic ground state [Fig.~\ref{magstr}(a)] is that the spiral rotation plane of the Ce moments lies within the plane defined by each Ce zigzag chain. In the spin Hamiltonian, this constraint can be encoded via bond-dependent anisotropy in the exchange matrices associated with $J_{\mathrm{out1}}$ and/or $J_{\mathrm{out2}}$. For instance, consider the $J_{\mathrm{out1}}$ bond connecting Ce sites at positions $x_i = (0.616, 0, 1/4)$ and $x_j = (0.384, 0, 3/4)$ (in Cartesian coordinates, where the first component is along the $a$-axis). The symmetry-allowed form of the $3 \times 3$ exchange matrix is:

\begin{align}
\hat{\mathcal{H}}_{ij} =
\mathbf{S}_i^T
\begin{pmatrix}
J_{xx} & 0 & J_{xz} \\
0 & J_{yy} & 0 \\
J_{xz} & 0 & J_{zz}
\end{pmatrix}
\mathbf{S}_j,
\label{Eq:matrix}
\end{align}

where $\mathbf{S}i$ is the spin operator at site $i$, and $T$ denotes transpose. The observed spiral plane can be naturally stabilized by introducing exchange anisotropy such that $|J_{xx}| = |J_{zz}| > |J_{yy}|$.

Once the incommensurate spiral along the $c$-axis is established by the interplay of $J_{\mathrm{out1}}$ and $J_{\mathrm{out2}}$, a sixfold-symmetric spin configuration [see Fig.~\ref{magstr}(c)] requires phase locking between spirals on the three Ce zigzag chains. Specifically, the in-plane components of Ce spins residing on the same layer must be related by a sixfold rotation about the $c$-axis, while their out-of-plane components should be ferromagnetically aligned. This phase synchronization can be achieved by introducing subdominant in-plane interactions $J_{\mathrm{in1}}$ and $J_{\mathrm{in2}}$, being antiferromagnetic for the in-plane components of the magnetic moments and ferromagnetic for the out-of-plane components ($J_{zz}<0<J_{xx,yy}$ for the in-plane interactions). The $J_{\mathrm{in1}}$ bonds form a corner-sharing triangle network equivalent to the kagome lattice geometry, while further including $J_{\mathrm{in2}}$ yields a full triangular lattice interaction network. Notably, both geometries classically stabilize 120$^\circ$ spin structures, i.e., the desired phase coherence among chains. In addition, the magnitudes of $J_{\mathrm{in1}}$ and $J_{\mathrm{in2}}$ likely remain weaker than $J_{\mathrm{out1}}$, as overly strong in-plane interactions would strongly distort the co-planar cycloidal structure established by $J_{\mathrm{out1}}$ and $J_{\mathrm{out2}}$, leading to deviations from the harmonic single-$q$ spiral observed in neutron diffraction. This component-dependent sign of the exchange interactions---together with the inferred anisotropy $|J_{xx}| = |J_{zz}| > |J_{yy}|$ of $J_{\mathrm{out}}$ interactions associated with the spiral rotation plane---highlights the crucial role of exchange anisotropy in stabilizing the collective magnetic behavior in this system~\cite{gauthier2024}. Such anisotropy is likely promoted by the spin–orbit-entangled nature of the Ce$^{3+}$ Kramers doublet.

\begin{figure}
\includegraphics[width=1\columnwidth]{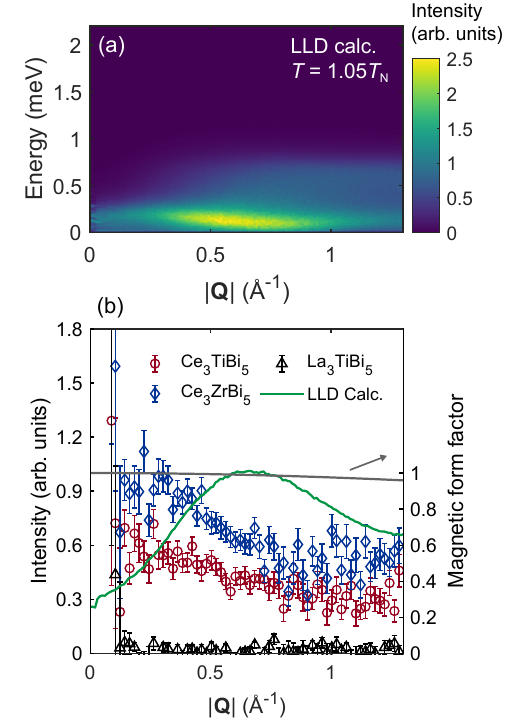} 
\caption{\label{sunny} Theoretical simulations of the collective excitation spectrum of Ce localized moments and comparison with experiment. (a) Energy- and momentum-resolved INS spectrum simulated using stochastic Landau–Lifshitz dynamics at $T = 1.05T_{\mathrm{N}}$, based on the phenomenological spin model described in the main text. (b) Comparison of the momentum-dependent spectral weight distribution between experimental data, simulation (solid green line), and the Ce magnetic form factor (solid grey line). For the simulations, the spectral weight was integrated over the entire positive energy transfer range. For the experimental data, the integration was performed over $0.25<E<0.625$\,meV.}
\end{figure}

With the minimal set of constraints to describe the magnetic structure---namely, $J_{\mathrm{out1}}<0$ with slight exchange anisotropy, $J_{\mathrm{out2}}=-0.725J_{\mathrm{out1}}$, and weak in-plane couplings $J_{\mathrm{in1}},\,J_{\mathrm{in2}}$ that are antiferromagnetic (ferromagnetic) between in-plane (out-of-plane) moments---we constructed a phenomenological spin model that captures the essential features of the ground state. This model was then used to estimate the energy- and momentum-resolved dynamical structure factor in the paramagnetic phase. For the simulation, we adopted $J_{\mathrm{out1}} = -0.42$\,meV, $J_{yy}=0.6J_{xx}$ for the $J_{\mathrm{out1}}$ interaction, and $J_{\mathrm{in1}} = J_{\mathrm{in2}} = -0.1 J_{\mathrm{out1}}$. We emphasize that these exchange parameters are not intended to represent quantitatively accurate exchange parameters, but rather to provide only a qualitative benchmark for the expected spectral profile of collective excitations from Ce$^{3+}$ local moments in Ce$_{3}$(Ti,Zr)Bi$_{5}$. Also, while our minimal model does not capture the opposite rotation directions of the two spirals within a single Ce zigzag chain [Fig.~\ref{magstr}(b)], this omission would not affect the overall momentum-space spin–spin correlation profile, as both configurations are characterized by the same magnetic propagation vector anyway. The simulations was conducted with a $24\times24\times24$ supercell of Ce$_{3}$(Ti,Zr)Bi$_{5}$ and a spin rescaling constant $\kappa =1.18$ (see Refs.~\cite{park2024_1,dahlbom2024} for more explanation of this parameter) at $T = 1.05 T_{\mathrm{N}}$.

As shown in Fig.~\ref{sunny}(a), the simulated powder-averaged inelastic neutron scattering (INS) spectrum at $T = 1.05 T_{\mathrm{N}}$ reveals a pronounced intensity centered around $|\mathbf{Q}| = 0.75$\,\AA$^{-1}$. This momentum corresponds to the magnetic Bragg peak at $(1, 0, 0.388)$ (r.l.u.), where strong spin-spin correlations are naturally expected near $T_{\mathrm{N}}$. Importantly, this Bragg peak corresponds to the lowest momentum transfer among all magnetic reflections derived from the magnetic structure in Fig.~\ref{magstr}. In contrast, the experimental data in Figs.~\ref{SEQpowder}(g) and (h) show that the diffuse magnetic signal is concentrated at much lower momentum transfers, around $|\mathbf{Q}| = 0.25$\,\AA$^{-1}$. This discrepancy is more clearly illustrated in Fig.~\ref{sunny}(b), where the calculated and observed spectra are directly compared. Such fundamentally different $|\mathbf{Q}|$ profiles strongly suggest that the low-energy excitations in Figs.~\ref{SEQpowder}(g)--(h) cannot be attributed primarily to collective modes of localized spins. Instead, they are more likely dominated by Kondo-related spin fluctuations, with only minor contributions from collective excitations of local moments embedded.

\begin{figure}
\includegraphics[scale=0.75]{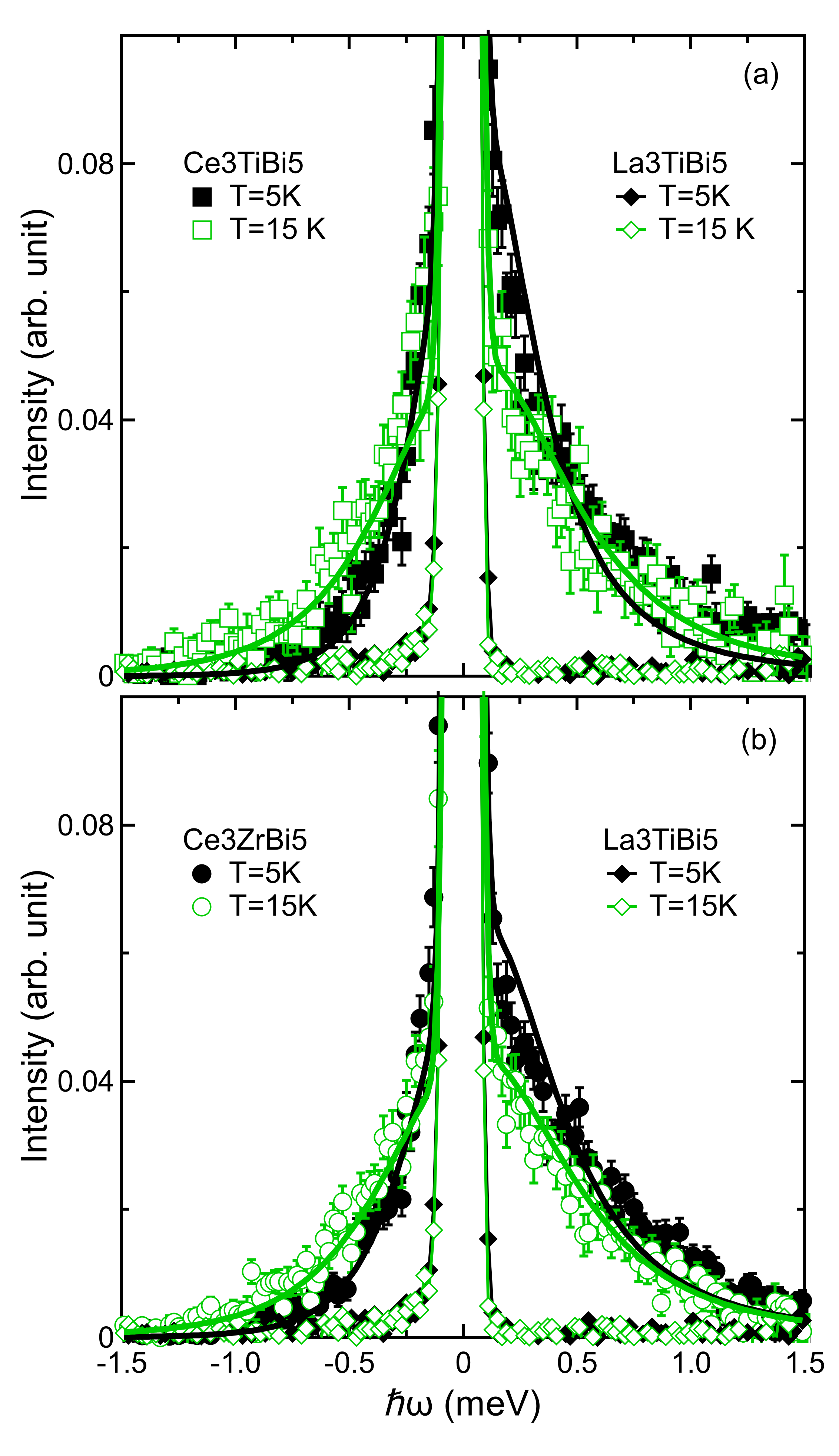} 
\caption{\label{linewidth} $T=5$~K and $T=15$~K measured scattering intensity as a function of energy transfer, $\hbar\omega$, for the $E_{\mathrm{i}}=4$~meV measurements integrated for wave-vector transfers between  $0.275 \leq Q \leq 0.625$ inverse Angstroms for (a) Ce$_3$TiBi$_5$ and (b) Ce$_3$ZrBi$_5$.  Measurements of La$_3$TiBi$_5$ are shown for comparison in each panel.  Error-bars for the La$_3$TiBi$_5$ data are not shown for clarity of the figure.  Solid lines are the respective global fits to the damped harmonic oscillator model described in the text. Data shown have had the measured empty can results subtracted from them and then normalized by the calculated nuclear scattering cross-section.}
\end{figure}

Building on the conclusion above, we analyze the energy dependence of the observed low-energy excitation spectrum, which provides a first-order estimate of the characteristic Kondo temperature. Figure~\ref{linewidth}(a) and (b) are the momentum-integrated scattering intensity as a function of tenergy transfer for $T=5$ and $T=15$\,K for both Ce compounds and the La-compound.  The additional spectral weight is clearly evident for both the Ce compounds compared to the non-magnetic analog.  A broad, asymmetric Lorentzian-like lineshape is evident in both Ce-based compounds. As a baseline, we fit the La-compound spectrum using a single Gaussian function with fixed width, representing the instrumental resolution and accounting for purely elastic scattering without Kondo-related fluctuations. The Ce-compound spectra are then modeled as the sum of this fixed-width Gaussian component and a damped harmonic oscillator (DHO) term, constrained by detailed balance, to capture the quasi-elastic response arising from fluctuating magnetic moments. The DHO response function takes the form:

\begin{align}
    I(\hbar\omega) = \frac{A}{1-\exp{(\frac{-\hbar\omega}{k_BT})}} & [\frac{\Gamma}{(\hbar\omega-\hbar\omega_0)^2+\Gamma^2}- \nonumber \\ &\frac{\Gamma}{(\hbar\omega+\hbar\omega_0)^2+\Gamma^2}],
\end{align}

where $A$ is an overall scale factor and $k_\mathrm{B}$ is Boltzmann's constant. We also include an overall constant background term to account for any residual background signal. We fit both temperatures for a given compound simultaneously to this parameterization of the scattering.  The La-compound spectrum yields a Gaussian full width at half maximum (FWHM) of 0.093\,meV; this value was fixed in subsequent fits to the Ce compounds.  We fix this value and perform a global fit separately to the Ce compounds. The results of the global fit, shown as solid lines in Fig.~\ref{linewidth}, show reasonable agreement with the data.  The value of $\hbar\omega_0$ was found to be 0.005 and 0.008\,meV for the Ce$_3$ZrBi$_5$ and Ce$_3$TiBi$_5$, respectively. The corresponding damping parameters were $\Gamma=0.43(6)$ and $\Gamma=0.72(5)$ for Ce$_3$TiBi$_5$ at $T=5$ and $T=15$\,K, respectively, and $\Gamma=0.53(7)$ and $\Gamma=0.72(6)$\,meV for Ce$_3$ZrBi$_5$. The mode energy and damping parameters are similar for both Ce compounds.  The uncertainties in the values for $\hbar\omega_0$ were on the order of 1\,meV. 

\comment{A Kondo temperature scale has been estimated from the fitted linewidth broadening $\Gamma$, by assuming a linear temperature dependence of it [$\Gamma(T)=\Gamma_0+aT$] and extrapolating to $T\to0$ to obtain $\Gamma_{0}$. This procedure yields $\Gamma_0=0.285(72)$\,meV for Ce$_3$TiBi$_5$ and $\Gamma_0=0.435(84)$\,meV for Ce$_3$ZrBi$_5$, corresponding to $T_{\mathrm{K}}\simeq \Gamma_0/k_B = 3.3(0.8)$\,K and $5.1(1.0)$\,K, respectively. These magnitudes---comparable to $T_{\mathrm{N}}=5$\,K for both compounds---indicate fairly strong Kondo hybridization competing with long-range order, well aligned with the heavy-fermion tendency inferred from thermodynamic response $\gamma=210\,\mathrm{mJ\,K^{-2}\,mol^{-1}}$~\cite{motoyama2018}.}

Finally, the momentum dependence of the low-energy excitations also merits more discussion. In both Ce$_3$TiBi$_5$ and Ce$_3$ZrBi$_5$, the intensity decays more rapidly with increasing $|\mathbf{Q}|$ than predicted by a generic magnetic form factor of Ce$^{3+}$ $4f^{1}$ (which merely changes across the $|\mathbf{Q}|$ range covered by our $E_{\mathrm{i}} = 4$\,meV measurement); see Fig.~\ref{sunny}(b). In a conventional single-ion Kondo scenario, the corresponding scattering intensity is expected to be momentum-independent aside from modulation by the magnetic form factor. This behavior may suggest a collective character of the fluctuations, which might not be surprising in the context of a Kondo lattice where hybridization effects could go beyond a single-ion picture. Investigating this likely Kondo-origin scattering and its anomalous momentum dependence remains an interesting direction for future work. In particular, cold-neutron spectroscopy with improved energy resolution, especially within the magnetically ordered phase ($T < 5$\,K), will be illuminating.

\section{Conclusion}
\comment{Our combined neutron diffraction and INS measurements establish Ce$_3$TiBi$_5$ and Ce$_3$ZrBi$_5$ as locally noncentrosymmetric, Ce-based Kondo-lattice antiferromagnets that exhibit unusual long-range magnetic order and pronounced hybridization effects. Neutron diffraction confirms the incommensurate cycloidal antiferromagnetic structure in Ce$_3$TiBi$_5$ and reveals a similar magnetic diffraction profile in Ce$_3$ZrBi$_5$, with ordered moments quite reduced relative to expectations based on single-ion $g$-factors. INS measurements resolve two well-separated crystal electric field excitations of the Ce$^{3+}$ $J = 5/2$ multiplet in both compounds. The corresponding ground-state doublet, as determined from the CEF analysis, is an admixture of $\ket{\pm 5/2}$, $\ket{\mp 3/2}$, and $\ket{\pm 1/2}$ states, not surprising given the low point symmetry of the Ce$^{3+}$ site. Among these, the $\ket{\pm 5/2}$ component is the most dominant.} 

\comment{At low energies, we observe a broad, quasielastic magnetic response with a bandwidth of order 1\,meV emerging above $T_{\mathrm N}$. Its qualitative disagreement with spin-dynamics calculations based on the ordered structure demonstrates that these fluctuations are not conventional local-moment excitations, but instead are dominated by Kondo-lattice dynamics. Consistent with this interpretation, an estimate of the Kondo temperature derived from the intrinsic quasielastic linewidth yields $T_{\mathrm{K}} \approx 3-5$\,K, comparable to $T_{\mathrm N}$, indicating strong competition between Kondo hybridization and magnetic exchange. This competition naturally accounts for both the pronounced ordered-moment suppression and the apparent magnetic ordering along the hard axis ($\parallel c$-axis), where the harmonic single-$q$ spiral involves significant out-of-plane spin components.}

\comment{Placed within the broader context of noncentrosymmetric Ce-based Kondo lattices, Ce$_3$TiBi$_5$ and Ce$_3$ZrBi$_5$ resemble systems such as CeRhSi$_3$, where the magnetic excitation spectrum is largely governed by Kondo-driven fluctuations~\cite{pasztorova2019}, rather than compounds like CePt$_3$Si that support more clearly defined collective modes of localized moments~\cite{Faak2009}. This points to sizable $c$–$f$ hybridization of the Ce $4f$ electrons in both materials. At the same time, both materials develop weak but well-defined long-range magnetic order. Thus, we conjecture that they are in the regime where Kondo screening and Ruderman–Kittel–Kasuya–Yosida (RKKY) interactions are closely competing.} 

\begin{acknowledgments}
This research was supported by the U.S. Department of Energy, Office of Science, Basic Energy Sciences, Materials Science and Engineering Division. This research used resources at the High Flux Isotope Reactor and Spallation Neutron Source, DOE Office of Science User Facilities operated by the Oak Ridge National Laboratory. The beam time was allocated to VERITAS on proposal number IPTS-32021.1, WAND$^2$ on proposal number IPTS-31482.1, HB-2A on proposal IPTS-30882.1, and SEQUOIA on proposal number IPTS-30893.1.
\end{acknowledgments}


\providecommand{\noopsort}[1]{}\providecommand{\singleletter}[1]{#1}%

\end{document}